\documentclass[epj,referee]{svjour}
\usepackage[dvips]{graphicx}
\begin{document}
\title
{Optimizing the photoassociation of cold atoms by use of chirped laser pulses}

\author{Eliane Luc-Koenig$^1$,  Mihaela Vatasescu$^{1,2}$, and Fran\c{c}oise Masnou-Seeuws$^1$ %
}                     
\institute{$^1$ Laboratoire Aim\'e Cotton, CNRS, B\^at. 505 Campus d'Orsay,
91405 Orsay Cedex,$^2$ Institute of Space Sciences, MG-23, RO-76911, 
Magurele-Bucharest, Romania. }
\date{Received: date / Revised version: date}
%
\abstract{Photoassociation of ultracold  atoms induced by chirped picosecond
pulses is analyzed in a non-perturbative treatment by following the wavepackets
dynamics on the ground and excited surfaces. The initial state is described by  a
Boltzmann distribution of continuum scattering states. The chosen example is
photoassociation of cesium atoms at  temperature T=54 $\mu K$ from  the $a^3
\Sigma_u^+(6s,6s)$ continuum  to bound levels in the external well of the
$0_g^-(6s+6p_{3/2})$ potential. We study how the modification of the pulse
characteristics (carrier frequency, duration, linear chirp rate and intensity) can
enhance the number of photoassociated molecules and suggest ways of optimizing
the production of stable molecules.}
\authorrunning{E. Luc-Koenig, M. Vatasescu, and F. Masnou-Seeuws}
\titlerunning{Cold atoms photoassociation using chirped laser pulses}
\maketitle

\section{Introduction}
The various routes leading to the formation of cold and ultracold molecules are actively explored \cite{masnou01}. Non-optical techniques like buffer gas cooling of molecules \cite{weinstein98} and Stark deceleration of polar molecules \cite{bethlem99,bethlem02} reach temperatures well below 1 K. Another route relies on optical techniques, laser fields being used to cool alkali atoms and to create excited molecules via the photoassociation reaction \cite{thorsheim87}; subsequently, these molecules are stabilized, by spontaneous emission or other radiative coupling, into bound vibrational levels of the ground electronic state  \cite{fioretti98,takekoshi98,nikolov99,nikolov00,gabbanini00,kerman03}. The translational temperatures thus reached are much lower ($T \le 20$ $\mu K$). Nevertheless, such stable molecules are produced in a superposition of vibrational levels, the most of them very excited. Then, bringing the molecules to the lower vibrational level (v=0) , thus making vibrationally cold molecules, is an important issue.

Up to now, most photoassociation experiments are using continuous lasers, but there are a few papers treating the photoassociation with pulsed lasers \cite{boesten96,gensemmer98,fatemi01}.  Theoretically, several time-dependent studies of photoassociation in the ultracold regime have been proposed \cite{mackholm94,vardi97,vala01,vatasescu01,eluc04}.

\begin{figure}
\resizebox{0.5\textwidth}{!}{%
\includegraphics{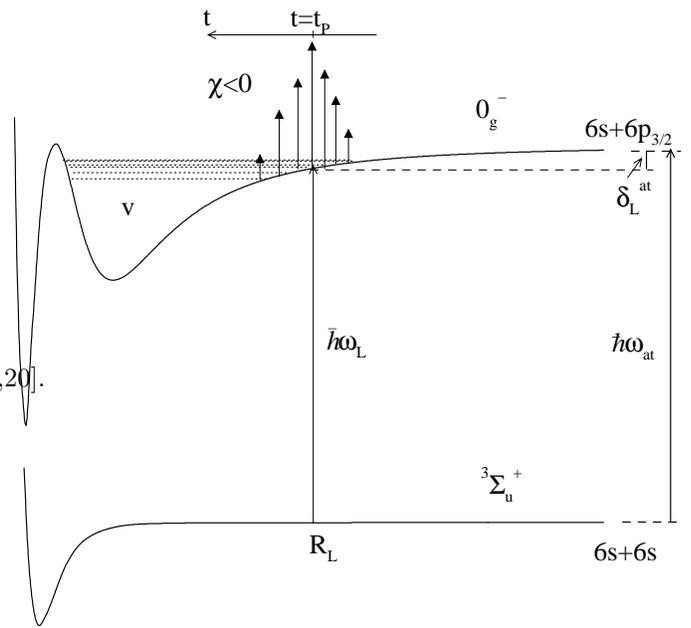}
}
\caption{Scheme of the photoassociation process with a negative chirped pulse considered in the present work, illustrated in the case of Cs$_2$. The potentials curves correspond to the ground triplet state a$^3\Sigma_u^+$(6s + 6s) and to 0$_g^-$ (6s + 6p$_{3/2}$) excited electronic state. In the present work, the energy of the initial continuum state is neglected in the definition of the resonance condition. The double well behaviour in the excited curve is a particular feature of the chosen symmetry. }
\label{fig:fig1}
\end{figure}

Our aim is to investigate the possibility to control cold molecules formation by use of chirped laser pulses. Indeed, it was shown that picosecond frequency-swept laser pulses produce more selective excitation and better population transfer than transform-limited pulses with the same bandwidth, duo to the mechanism of population inversion by adiabatic sweeping \cite{baum85,melinger91,melinger92,goswami02}. \\
 Photoassociation of cold atoms with a chirped laser pulse was first explored theoretically by Vala et al. \cite{vala01}, using a gaussian packet centered at large interatomic distance ($R \approx 200\  a_0$) as initial state in the collision of two cesium atoms at the temperature $T=200$ $\mu K$, and showing that a picosecond pulse can achieve a total transfer of population under adiabatic following conditions proposed by Cao, Bardeen, and Wilson \cite{cao98,cao00}. Nevertheless, such an approach cannot work at collision energies close to threshold, and if we are interested in the dynamics at smaller distances (which it is the case if the aim is to produce cold molecules in low vibrational levels), because a gaussian wavepacket does not adress the actual shape of the initial continuum state. Indeed, in an ultracold collision of two atoms, the  kinetic energy $k_B T$ becomes easily smaller than the interaction potential, and the correct representation of the initial continuum state is a stationary collisional state. Moreover, at a given collision energy, the behaviour of a continuum state varies a lot  with the interatomic distance, and then the distance range where the excitation takes place (for example, $R_L$ in Fig. \ref{fig:fig1}) is of great importance for the photoassociation yield. Therefore, a theoretical treatment of the dynamics at very low temperatures needs a quite precise representation of the initial state.  

In a previous theoretical paper \cite{eluc04}, we have considered the photoassociation of two cold cesium atoms with a picosecond chirped pulse that excites several vibrational levels in the $0_g^-(6s+6p_{3/2})$ external well. This reaction has been widely studied experimentally with continuous lasers, so that the relevant potential curves are well known. The time-dependent Schr\"odinger equation corresponding to the two-electronic coupled channels ($a^3 \Sigma_u^+(6s+6s)$ and $0_g^-(6s+6p_{3/2})$) was solved using a propagation method \cite{kosloff94,kosloff96}, and, due to a Mapped Sine Grid representation \cite{willner04}, it was possible to take as the initial state in the photoassociation a stationary collision state corresponding to a temperature of about 50 $\mu K$. The chirped pulse has been designed in view of creating a vibrational wavepacket in the excited state, which, after the pulse, moves to shorter internuclear distances and is focussing at the barrier of the double well potential, thus preparing a good initial state for the stabilization process  into low vibrational levels of the $a^3 \Sigma_u^+(6s+6s)$ potential. Besides its property of focalization, the specificity of the picosecond  chirped pulse analyzed in Ref. \cite{eluc04}  is that it leads to a strong transfer of population inside a spatially limited ``photoassociation window'', and no tranfer outside. Finally, the population transfer to the last vibrational levels of the ground $a^3 \Sigma_u^+(6s+6s)$ state is significant, making stable molecules. We have shown that these results can be interpreted in the framework of a two-state model and in the impulsive limit \cite{banin94}, as an adiabatic population inversion taking place in the ``resonance window'' swept by the pulse. \\
In the present work, we consider the same photoassociation reaction, and analyse how the results are modified by acting with various pulses,  in order to determine how  characteristics like duration, chirp rate and energy  influence the photoassociation yield. The results obtained using pulses of different durations (ps, tens and hundreds of picoseconds) and energies,  can then be used to explore various regimes of population transfer, nonadiabatic effects, and the eventual limits of the impulsive approximation. Moreover, as our calculation are performed using a very large spatial grid (thousands of $a_0$), the results account for the threshold effects intervening in this continuum-bound transfer of population  taking place at very low energies and implying large ranges of distance in the transition (as the initial state is a delocalized collisional continuum). \\
The exploratory results thus obtained will be used in the calculation of the effects of a second pulse, which can be either one identical with the first (to estimate the effect of the repetition rate), either a different one (chosen for the stabilization of the packet focalized in the $0_g^-$ channel to  vibrational levels of the $a^3 \Sigma_u^+$ potential). Obviously, this non-perturbative model allows to treat laser pulses of strong intensities and non-resonant effects.

The  photoassociation reaction studied here is between two cold cesium atoms colliding in the ground state potential $a^3 \Sigma_u^+(6s+6s)$, at a temperature $T=54$ $\mu K$, which are excited by a pulse laser to form a molecule in a superposition of vibrational levels $\{ v \}$  of the  electronic potential $0_g^-(6s+6p_{3/2})$. For  a rotational  quantum number J=0, it can be represented  as:
\begin{eqnarray}
Cs(6s^{2}S_{1/2})+Cs(6s^{2}S_{1/2})+\hbar\omega(t) \nonumber \\
 \rightarrow Cs_{2}(0_g^-(6s^{2}S_{1/2}+6p^{2}P_{3/2}); \{ v \}, J=0),
\label{eq:photo}
\end{eqnarray}
We consider the excitation by a chirped laser pulse of gaussian envelope,  having a time-dependent frequency $\omega(t)/2 \pi$ which varies linearly around the central frequency $\omega_L/2\pi$ reached at $t=t_P$, and  red-detuned by  $\delta^{at}_{L}$  relative to the $D_2$  atomic resonance line:
\begin{equation}
\hbar\omega(t_P)=\hbar\omega_L=\hbar\omega_{at}-\delta^{at}_{L}.
\label{eq:deltatL}
\end{equation}
$\hbar \omega_{at}$ is the energy of the atomic transition 6s $\to$ 6p$_{3/2}$.
The central frequency of the pulse, $\omega_L/2\pi$, or equivalently the detuning $\delta^{at}_{L}$,  determines the crossing point $R_L$ of the two electronic potentials dressed by the photon with the energy $\hbar\omega_L$ (see Fig. \ref{fig:fig1}). In the present calculations  the  detuning is fixed at $\delta_L^{at}$=2.656 cm$^{-1}$, corresponding to $R_L=93.7 \ a_0$ and to 
resonant excitation at $t=t_P$ of the level $v_0$=98 in the external well of the 0$_g^-(6s+6p_{3/2})$ potential. Then, taking the origin of energy as being the dissociation limit $6s+6p_{3/2}$ of the 0$_g^-(6s+6p_{3/2})$ potential, the binding energy of the $v_0$ level is $E_{v_0}=\delta_L^{at}$. In this paper the origin of energy is chosen at the dissociation limit $(6s+6s)$ of the $a^3 \Sigma_u^+$ potential, the energy of the 0$_g^-$ $v_0$ level being then equal to 0. \\
The potentials used in our calculation have been described in a precedent paper \cite{eluc04}. The outer well of the 0$_g^-(6s+6p_{3/2})$ excited potential was fitted to photoassociation spectra  by Amiot {\it et al} \cite{amiot02} and matched to  {\it ab initio} calculations at short and intermediate range \cite{pellegrini03} . 
The  $a^3\Sigma_u^+(6s,6s)$ potential has been chosen in order to reproduce correctly the scattering length  $L \approx 525 \ a_0$ and the asymptotic behaviour $-C_6/R^6$ with $C_6$= 6828 au \cite{amiot02} (the short range part extracted from the Ref. \cite{spies89} being slightly modified for that purpose).

The paper is organized as follows: in Section 2 we discuss the representation of the initial state in the photoassociation dynamics. Section 3 describes the numerical methods used to perform  time-dependent propagation on very large spatial grids. Section 4 shows  the two-channel model used for the photoassociation, the basic formulae for gaussian pulses  with linear chirp, the choice of the pulse parameters, and treats  specific  features of the excitation with a chirped pulse, as the  energy range excited resonantly or not resonantly, and the condition for an ``adiabaticity window'' during the pulse duration. Section 5 compares the results obtained with pulses differing as  duration, energy, and chirp. In Section 6 we show the calculation of the photoassociation probability from a thermal average over the incident kinetic energies, and of the number of molecules photoassociated per pump pulse. Section 7 presents a discussion on the possible ways for the optimization of the process. Section 8 is the conclusion. The article has three appendix, containing: Appendix A (Section 9) - the energy normalization of the ground state continuum wavefunction calculated in a box;  Appendix B (Section 10) - the calculation of the chirp rate in the time domain for focussing the excited vibrational wavepacket at the outer turning point;  Appendix C (Section 11) - the overlap of the initial continuum with the 0$_g^-$ vibrational wavefunctions.

\section{Representation of the initial state in the photoassociation dynamics}

We shall briefly review here the description of the initial state of the photoassociation process, as a thermal equilibrium state in a  gas of cold alkali atoms, lying in a trap of infinite dimension with the laser turned off (then not interacting with the electromagnetic field), and being initially in their ($ns$) ground state. Supposing low atomic densities, which allows to consider the  binary events  isolated enough, the N-body ($N \geq 3$) interactions can be ignored, and the Hamiltonian will be reduced to a summation of non-interacting pair Hamiltonians, treated in the Born-Oppenheimer approximation \cite{pillet97}. Then, for a colliding pair of atoms, the molecular Hamiltonian $H_M$ of the relative motion can be separated from the center of mass motion which does not play any role in photoassociation. Assuming thermal equilibrium at temperature T, the initial state is a statistical mixture of eigenstates corresponding to $H_M$, being described by the density operator:
\begin{equation}
\hat{\rho} = \frac{e^{-\beta \hat{H}_M}}{Z}, \ \beta =1/(k_B T)
\label{eq:boltz}
\end{equation}
with $Z=Tr\{e^{-\beta \hat{H}_M}\}$  the partition function ($k_B$ being the Boltzmann constant).  For a gas of low density composed of non-interacting pairs of atoms in a volume V, the partition function is \cite{bruhat}:
\begin{eqnarray}
Z=Q(T)V, \  \  Q(T)=\frac{(2\pi \mu k_BT)^{3/2}}{h^3},
\label{eq:partitionz}
\end{eqnarray}
$\mu$ being the reduced mass of the diatom. 

The density operator (\ref{eq:boltz}) can be decomposed into partial waves $l$ describing the relative rotational angular momenta: 
\begin{eqnarray}
\hat{\rho} = \frac{1}{Z}\sum_{l,g} { (2l+1)  \hat{\rho}^g_l},  \  \hat{\rho}^g_l = e^{-\beta  \hat{H}^g_l}  
\end{eqnarray}
where the molecular Hamiltonian $\hat{ H^g_l} = \frac{\hat{P}_R^2}{2 \mu}+V_{g}(R)+\frac{\hbar^2 l(l+1)}{2 \mu R^2}$. $R$ designs the interatomic coordinate, $\frac{\hat{P}_R^2}{2 \mu}=\hat{T}$ is the kinetic energy operator, and $V_{g}(R)$  is the potential of the ground electronic state (neglecting the hyperfine structure,
$g$ represents one of the states $X^1\Sigma_g^+(ns+ns)$ or $a^3\Sigma_u^+(ns+ns)$).
We shall restrict our study at  $l=0$ ($s$ wave), which is a good approximation for low energy collisions,  and show here calculations for cesium photoassociation, taking $a^3\Sigma_u^+(6s+6s)$ as electronic  ground state. Taking the dissociation limit $(6s+6s)$ of the $a^3 \Sigma_u^+$ potential as origin of energy,  the initial state of the photoassociation process will be described by a density operator $\hat{\rho}_s$:
\begin{equation}
\hat{\rho}_s=\frac{1}{Z} \int_0^{\infty} dE e^{-\beta E}|E> <E|
\label{eq:averen}
\end{equation}
where $|E> \equiv |^3\Sigma_u^+,  l=0, E>$ are continuum  eigenstates (normalized per unit energy) of the Hamiltonian $\hat{H}_{l=0}$, of  energy E and  corresponding to the relative motion in the electronic potential $^3\Sigma_u^+$ and to $l=0$ (s -wave).

In the following we shall discuss the representation of the initial state in the study of the photoassociation dynamics by time-dependent propagation. An evaluation of the initial density matrix by gaussian wavepackets can be valid  for sufficiently low laser detunings $\delta^{at}_{L}$ (very large distance R) and sufficiently high temperatures T \cite{vala01,vatasescu01}. On the contrary, at very low temperatures and for the photoassociation in more profound vibrational states of the excited molecular potential, the potential energy $V_g(R)$ becomes significant relative to the kinetic energy term, and the calculation needs to account for the nodal structure of the ground state. In this case, the evaluation of the initial density matrix by stationary collision states is required. 
In this paper we present calculations corresponding to this second case, showing that such delocalized continuum states can be represented numerically and used in time dependent  propagation.

\subsection{Evaluation of the initial density matrix by gaussian wavepackets}
For a sufficiently small detuning $\delta_L^{at}$, the crossing point $R_L$ of the electronic potentials dressed by $\hbar\omega_L$  occurs at large distance. In the case of the excitation with a continuous laser, and for a wavepacket localized around $R_L$, the crossing point $R_L$ completely determines the R-domain governing the photoassociation process. For the photoassociation with a pulse, other factors strongly influence the distance range which can be excited: the duration and the intensity of the pulse,  its chirp rate (see Sec. \ref{sec:enrange}), and the dynamics depends on the spatial localization or delocalization of the initial state.

For sufficiently high temperatures T, one can expect that the kinetic energy dominates the interaction  potential $V_g(R) \sim -\frac{C_6}{R^6}$ at large R distances:
\begin{equation}
k_BT \gg \frac{C_6}{R^6},
\label{eq:gaussap}
\end{equation}
resulting in a relatively uniform R-dependence for the density matrix. If the inequality (\ref{eq:gaussap}) is valid, the density operator can be decomposed in a set of incoming gaussian wavepackets with mean velocity $\bar{v}_{0}$ equal to the most probable velocity of a Maxwell-Boltzmann distribution (then the corresponding momentum $p_0=\mu \bar{v}_0=\sqrt{2\mu k_BT}=\sqrt{<p_0^2>}$, and the momentum spread is defined as $\Delta p_R =\frac{p_0}{2}$). The width $\Delta R$ of the radial density distribution is related to the momentum spread for the relative motion by $\Delta R \Delta p_R \approx \frac{\hbar}{2}$, then depending on  the temperature \cite{vala01,vatasescu01}.

In the case of the  $a^3\Sigma_u^+(6s+6s)$ state of the Cs$_2$, for the temperature $T \approx 50$ $\mu K$ studied in the present paper, the relation (\ref{eq:gaussap}) is valid for $ R \gg 185$ a$_0$, which corresponds to  resonant excitation of vibrational states with $v \gg 130$ in the $0_g^-(6s+6p_{3/2})$ external well,  or to a laser detuning $\delta_L^{at} \ll 0.35$ cm$^{-1}$.   At  $T \approx 50$ $\mu K$ (a collision with relative velocity $\bar{v}_0 \approx 10$ cm/s), the width corresponding to a radial wavepacket is very large: $\Delta R \approx 160$ a$_0$ (in cold collisions the large de Broglie wavelength describing the relative motion is the sign of the wavefunction delocalization). 

Gaussian wavepackets have been succesfully used in the study of photoassociation \cite{vala01} at a large distance  $R_L = 200$ a$_0$ and a temperature  $T = 200$ $\mu K$,  for a pulse duration (75 ps) sufficiently small to avoid observable motion on the ground surface. 

On the opposite, in a precedent paper \cite{vatasescu01} studying the photoassociation with a continuous laser in the electronic state $1_g(6s+6p_{3/2})$ of Cs$_2$, at $R_L = 90$ a$_0$ and for a temperature $T = 125$ $\mu K$, we have shown that a gaussian wavepacket having a large radial width and a small momentum $p_0$ is spreading more rapidly than it is moving; being quite difficult to gain insight on the relevant dynamics by using it, a possible choice is to consider a much smaller spatial width, implying a bigger momentum width. Components with  much bigger momenta are then introduced  in the representation of the initial state, falsifying the dynamics in the corresponding channel. 

Therefore the representation of the initial state by gaussian wavepackets cannot be used at very low temperatures if we  are interested in the real dynamics at small distances and  in the initial molecular channel, because it hardly adresses properly the physics at distances different from R$_L$. 

\subsection{Evaluation of the initial density matrix by stationary collisional eigenstates}
\begin{figure}
\resizebox{0.5\textwidth}{!}{%
\includegraphics{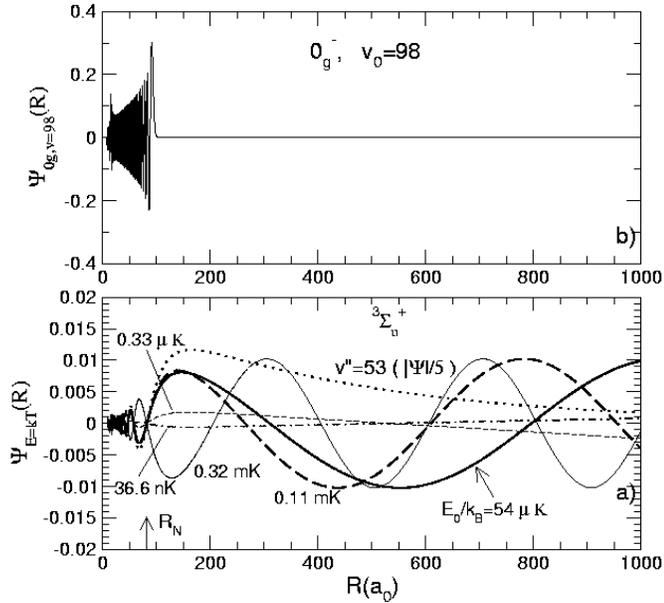}
}
\caption{a) Stationary continuum states in the $a^3\Sigma_u^+(6s+6s)$ potential, corresponding to energies $E=k_B T$, which are calculated on a grid of length $L_R=19250$ a$_0$ and normalized at 1 in the box. $R_N$ is the last node common to the collisional wavefunctions in the studied domain of energies. With the continuum thick line is represented the continuum wavefunction presently chosen as initial state in photoassocciation at the temperature $E_0/k_B =54$ $\mu K$. Other continuum wavefunctions obtained in the box are represented, corresponding to the following values of $E/k_B$: 36.6 nK (dot-dashed line), 0.33  $\mu K$ (short-dashed line), 0.11 mK (long-dashed line), and 0.32 mK (thin continuum line).  The last bound state $v''=53$ of the triplet potential is also shown (with the wavefunction diminished by a factor of 5); b) The wavefunction corresponding to the level $v_0=98$ in the outer well of the 0$_g^-(6s+6p_{3/2})$ potential, which is excited at $t=t_P$ ($E_{v_0}=-2.656$ cm$^{-1}$ under the dissociation limit $6s+6p_{3/2}$) .}
\label{fig:art2-init}
\end{figure}

At very low temperatures, the kinetic energy $k_BT$ can easily become smaller than the potential energy  $\frac{C_6}{R^6}$, even for large values of R.  Then it is absolutely necessary to account for the nodal structure of the ground state wavefunction, which is responsible for the intensity minima observed in the photoassociation spectra. This is the case for the studied example of photoassociation in the $0_g^-(6s+6p_{3/2})$ channel in Cs$_2$, at a detuning $\delta_L^{at} \approx 2.4$ cm$^{-1}$, $R_L\approx 94$ a$_0$, and $T \approx 50$ $\mu K$.   Indeed, in the ground state $a^3\Sigma_u^+(6s+6s)$,   all the stationary wavefunctions  involved in the presently studied energy range have the same nodal structure until $R_N = 82.3$ a$_0$ (the position of the last common node), differing only by a normalization factor (see Fig. \ref{fig:art2-init}a) ). For photoassociation at large distance $R_L \gg R_N$ the distribution of the continuum states is uniform, but if one has to account for a range of distances  such as $R_L \sim R_N$ it is necessary to describe the initial density matrix as an incoherent average over a thermal distribution of stationary energy-normalized continuum wavefunctions $|E>$, as in Eq. (\ref{eq:averen}). Each  incident component will contribute incoherently to the density probability $| \Psi_{0_g^-}(R,t) |^{2}$ transferred at the time t in the $0_g^-$ surface, with a weight equal to $\frac{e^{-\beta E}}{Z}$.

\subsection{Representation of the energy-normalized initial continuum state on a spatial grid.}
In our method, the eigenstates of the Hamiltonian associated with the ground electronic channel are calculated through the Sine Grid Representation \cite{willner04,kosloff96}, in a large box of radius $L_R$ which introduces a discretization of the continuum (only continuum states having a node at the boundary of the box are obtained) and supplies wavefunctions normalized to unity in the box. We shall label  $E_n>0$ the energies of these states and  $\phi_{E_n}(R)$ the corresponding wavefunctions having $n$ nodes in the range $R_N<R<L_R$ ($n \ge 0$ integer). The energy resolution in a box of length $L_R$ is determined by the energy difference $\delta E$ corresponding to neighbouring eigenstates of the box:
\begin{equation}
\delta E \mid_{E_n}=\frac{\partial E}{\partial n}\mid_{E=E_n} = \frac{\hbar \pi}{L_R}\sqrt{\frac{2E_n}{\mu}}=\frac{\hbar^2 \pi^2}{\mu L_R^2}n
\end{equation}
Then a large box is necessary to have a sufficiently high energy resolution (small $\delta E$), able to represent low energies corresponding to the studied temperature $T \approx 50$ $\mu K$. For the present calculations we use a box with the width $L_R= 19250$ a$_0$, in which the continuum state having the energy $E_n=k_B T$, with $T = 54$ $\mu K$, is the 38th state in the corresponding discretized continuum. This leads  to a resolution at threshold  $\frac{\delta E}{k_B}=30$  nK and
 $\frac{1}{k_B} \frac{\partial E}{\partial n}\mid_{E/k_B=54\mu K }=2.6$ $\mu K$ for the continuum description around 54 $\mu K$. Such a large box allows to account correctly for the threshold behaviour in the presently studied cold collision process and to perform explicitely an average on the thermal energy distribution. 

For an energy resolution $\delta E$, the longest relevant time scale which can be studied in the problem is  $\tau_{max}= \hbar /\delta E$. Any energy spacing smaller than 
$\delta E$  is not resolvable in the box of length $L_R$ (for example, an eigenstate characterized by a vibrational period bigger than $\tau_{max}$ cannot be distinguished from the continuum positive energy states in the box  determined by this $L_R$). In the present case $\tau_{max}$ is $2.9$ $\mu s$, being larger than the radiative lifetime in the $0_g^-$ state, $\tau_{rad}\approx 15$ ns; but the spontaneous emission is not introduced in the present problem, where the dynamics is followed for times smaller than $\tau_{rad}$.

The representation of the initial density matrix by stationnary continuum states needs a normalization per unit energy of the continuum wavefunctions.  The basic formulae are shown in the Appendix A: one shows that the relation between the wavefunctions $\Psi_{E_n}(R)$ normalized per unit energy and the  wavefunctions $\phi_{E_n}(R)$ normalized to unity in the box is the following:
\begin{equation}
\Psi_{E=E_n}(R)=\lbrack \frac{\partial E}{\partial n}\mid_{E=E_n}\rbrack^{-1/2}\phi_{E_n}(R)
\end{equation} 
 In the presently Grid representation the density of states is calculated explicitely from the energy spacing between two neighbouring states of the discretized continuum:
 \begin{equation}
\frac{\partial E}{\partial n}\mid_{E=E_n}= E_{n+1}- E_n
\end{equation}      

\section{Numerical methods}

The dynamics of the photoassociation process is performed by solving numerically the time-dependent Schr\"odinger equation for the ground $V_g(R)$  and excited $V_e(R)$ potentials  coupled by the electromagnetic field, and propagating the wavepackets on the ground and excited surfaces. The cold collision of the two atoms on the ground surface $V_g(R)$ is generally characterized by a huge local de Broglie wavelength $\Lambda(E,R)$, as the relative momentum $p_R$ is very low:
\begin{eqnarray}
\Lambda(E,R)=\frac{h}{\sqrt{2\mu [E-V_g(R)] }}=\frac {2\pi \hbar}{p_R}
\label{eq:debroglie} 
\end{eqnarray}
As we have shown in the  precedent section, we choose as initial condition a stationary continuum  state of the ground surface of very low energy corresponding to a temperature $T = 54$ $\mu K$. A grid of very large extension is needed to represent  correctly such an initial state and the last bound states of the ground potential which are populated during the photoassociation process, and whose wavefunctions extend at large interatomic distances. We consider a grid extending from $L_0$ (a distance slightly smaller than the repulsive walls in the potentials, here $L_0$=8 a$_0$ ) to $L_R= 19250$ a$_0$ in the present paper. This very large grid allows to  represent correctly the wavefunctions implied in the process and to account for the threshold behaviours at low energies in the realistic potentials. 

\subsection{Spatial representation of the wavefunctions}
The radial dependence of the wavepackets $\Psi_{e,g}(R,t)$ propagating on both surfaces is represented using Mapped Grid Methods \cite{slava99,willner04}. The mapping works with a change of variable (from R to the adaptive coordinate $x$)  made as to take into account  the variation of the local de Broglie wavelength $\Lambda(E,R)$ (see Eq. (\ref{eq:debroglie})) as a function of the internuclear distance R. Namely, the Jacobian of the transformation is chosen proportional to the local de Broglie wavelength: $J(x)=dR/dx= \beta \Lambda(E_{max},R)$, where  $E_{max}$ is the maximum energy involved in the problem. This leads to the implementation of a large spatial grid  by using only a  small number of points. For example, in our calculations we use N=1023 points for a spatial extension $L_R=19250$ a$_0$, describing uniformely regions of space where $\Lambda(E_{max},R)$ varies by several orders of magnitude. 
The use of an enveloping potential $V_{env}(R)$ (equal or deeper than the crossing $V_e$ and $V_g$ potentials dressed by the field)  allows the definition of a common Jacobian and therefore a single $x$-grid to describe both surfaces. In the new coordinate $x$, the kinetic energy operator expresses simply as product of operators $J(x)^{-1/2}$ and $d/dx$, without introducing derivatives of $J(x)$  \cite{slava99,willner04}. The potential energy operator reduces simply to $V(x)$. 

A collocation method \cite{slava99,willner04} is used to define the representation of the wavefunctions in the N points $x_i$ of the grid $(x)$. This supposes that any wavefunction $\bar{\Psi}(x)$ is expanded on a set  of N basis functions. Instead of the usual plane wave expansion (Fourier expansion), we use the Sine expansion recently introduced by Willner et al. \cite{willner04}. All basis sine functions have nodes at the boundaries $L_0$ and $L_R$ of the grid, which, by choosing a sufficiently small  $\beta$-value \cite{willner04},  permits the suppression of the so-called ``ghost'' levels appearing in the solution of the stationary Schr\"odinger equation and then susceptible to falsify the dynamics of the system.  We have verified that a  value  $\beta=0.52$ is necessary for avoiding the appearance of ``ghost'' levels. 

To determine the stationary eigenstates of the ground or  excited potential (which are used either for the selection of the initial state, or to analyze the wavepackets evolution in terms of their decompositions on stationary states) an auxiliary cosine basis set is introduced, allowing to evaluate analytically the first order derivatives $d/dx$ involved in the kinetic operator \cite{willner04}.

In time propagation calculations one has to multiply repeatedly the initial state wavefunction by 
 the Hamiltonian matrix. Therefore one has to express the kinetic energy operator in terms of discrete sine and cosine Fourier transformations \cite{borisov01}, for which we use the efficient fast Fourier algorithms. This imposes $N=2^p-1$ ($p$ integer) for the number of grid points (N=1023 for the present calculation).

Lastly, in time propagation calculation the initial state is normalized to unit in the box of size  $L_R$. Therefore the populations $P_{0_g^-}(t)$ or $P_{^3\Sigma_u}(t)$ calculated at a time t are depending on the value chosen for $L_R$. Nevertheless, due to the normalization condition $P_{0_g^-}(t)+P_{^3\Sigma_u}(t)=1$, which is verified for every $t$, the calculations give directly the relative population on each surface.

\subsection{Time Evolution}
The time-dependent Schr\"odinger equation for the two coupled channels (see Eq. (\ref{eq:cplfin})) is solved by propagating the initial wavefunction using a Chebychev expansion of the evolution operator  $\exp[-i \mathbf{\hat H}t/\hbar]$  \cite{kosloff94,kosloff96}. Propagation is performed in discrete steps with a time increase $\Delta t$  much shorter  than the characteristic times of the problem (pulse duration, vibrational periods, Rabi periods).  In the present study of photoassociation with chirped pulses the Hamiltonian $H(t)$ is explicitely time-dependent, and a discrete description of $H(t)$ is introduced. During the step corresponding to time propagation from $t_1$ to $t_2=t_1 + \Delta t$, the Hamiltonian is supposed to be time-independent and chosen as $H(t_m)$, where 
$t_m=(t_1 + t_2)/2$. Such a procedure introduces errors of the order of magnitude $(\Delta t)^3$.
Since presently the initial state is a stationnary collisional eigenstate in the $^3\Sigma_u^+$ potential, instead of a gaussian wavepacket, the dynamics results only from coupling with the laser pulse, but not from the evolution motion characteristic for a nonstationary state. Moreover, the dynamics is studied within a very large box. Consequently, even after a very long propagation duration of 15 ns (the order of magnitude of the spontaneous emission time for the vibrational levels excited in the $0_g^-$ state), the wavepackets dynamics in the range of distances relevant for our problem (in which the hyperfine structure is neglected) is not influenced by the external boundary of the box.
 Therefore there is no reflection of the wavepackets at $L_R$ and then it is not necessary to define outgoing wave boundary conditions for the wavepackets (either by transferring the outgoing part to another grid \cite{heather87} or by introducing an imaginary absorbing potential \cite{kosloff86}).

In the present problem, using a grid with N=1023 points, which for  $\beta=0.52$ gives a box of extension $L_R=19250$ a$_0$, the energy resolution at the $^3\Sigma_u^+$ dissociation threshold is determined by $\delta E/k_B \approx 100$ nK. For a grid with only 511 points, the extension is reduced at $L_R=2450$ a$_0$ (about 7.8 times smaller than $L_R$) and the energy resolution is strongly diminished, as $\delta E/k_B$ $\approx 6$ $\mu$K. Since we study the photoassociation taking an initial stationary state corresponding to T$\approx 50$ $\mu$K, a grid of large extension is needed to attain the necessary resolution in the representation of the threshold processes. As we have shown before, this is possible due to the mapping procedure, which drastically reduces the number of grid points, and due to the implementation of the Sine Basis representation, which eliminates the participation of the ``ghost levels'' in the dynamics.

For the step $\Delta t \approx 0.05$ ps used in our time propagation, $\mathbf{\hat H}\Psi $ is calculated 112 times during each interval  $\Delta t$. Therefore the implementation of the fast Fourier transformations to calculate ${\hat T}\Psi$ avoids prohibitive calculation times, allowing in principle  to study the evolution of the wavepackets during time durations as long as 10 ns (which corresponds to a repetition rate of $10^8$ Hz for the photoassociating laser pulse). Obviously, such long evolution times need to introduce in the problem other physical processes, as the spontaneous emission ($\tau_{rad}=15$ ns), which is not taken into account for the shorter times of our present calculations.

\section{Two-channel model for the photoassociation with a gaussian chirped pulse}

\subsection{Basic formulae for gaussian pulses with linear chirp}
We shall briefly review the main relations concerning pulses with gaussian envelope and linear chirp \cite{cao98,cao00,eluc04}.  The electric field has the amplitude ${\cal {E}}_0$, a gaussian envelope $f(t)$, the carrier frequency  $\omega_L/2\pi$, and a phase  $\varphi(t)$:
\begin{equation}
{\cal{E}}(t)={\cal {E}}_0 f(t) \cos [\omega_Lt + \varphi(t)],
\label{eq:field}
\end{equation}
The gaussian envelope $f(t)$ is maximum at $t=t_P$ and has the full width at  half maximum (FWHM) equal to $\sqrt{2}\tau_C$: 
\begin{equation}
f(t)=\sqrt{\frac{\tau_L}{\tau_C}} \exp [-2 \ln 2 (\frac{t-t_P}{\tau_C})^2],
\label{eq:env}
\end{equation}
The duration $\tau_L$ characterizes the spectral width $\delta \omega$= $4\ln 2/\tau_L$ in the frequency domain: indeed,  $\tilde{\cal E}(\omega)$, which is the Fourier transform of ${\cal {E}}(t)$,  displays a gaussian profile with FWHM equal to $\sqrt{2}\delta \omega$.

The ratio $\frac{\tau_C}{\tau_L} \ge 1 $ characterizes the chirp: the value $\tau_C = \tau_L$ corresponds to an unchirped transform-limited pulse with duration  $\tau_L$, which is streched to $\tau_C$ by chirping.   The linear chirp is determined by the chirp rates: 
$\chi$ in the time domain (calculated as the second  derivative of $\varphi(t)$, the phase of the field 
${\cal {E}}(t)$), and $\Phi^{\prime\prime}$ in the frequency domain (the second  derivative of the phase of $\tilde{\cal E}(\omega)$):
\begin{eqnarray}
\frac{\tau_C}{\tau_L} =  \sqrt{1+(4\ln2)^2\frac{(\Phi^{\prime\prime})^2}{\tau_L^4}} =
\sqrt{1+ \frac{\chi^2 \tau_C^4 }{(4\ln2)^2}}
\label{eq:stretch}
\end{eqnarray}
The choice of $\tau_L$ and  $\chi$ imposes the pulse duration $\tau_C$ (see Eq. (\ref{eq:stretch})), and then the chirp rate  $\Phi^{\prime\prime}$ in the frequency domain:
\begin{equation}
\Phi^{\prime\prime}=\chi \frac{\tau_C^2 \tau_L^2}{(4 \ln 2)^2} 
\label{eq:chirparam}
\end{equation}
Among the four parameters $\tau_C$, $\tau_L$,  $\chi$ and $\Phi^{\prime\prime}$, only two are independent (see Eqs. \ref{eq:stretch} and \ref{eq:chirparam}). Another independent parameter is the coupling $W_L$. Then, the choice of three independent parameters will determine the nature of the transfer between the two channels, adiabatic or non-adiabatic, as it will be shown in Sec. \ref{sec:adiabwind}.

Chirping a pulse increases its duration and decreases its maximum amplitude: ${\cal {E}}_M={\cal {E}}_0 \sqrt{\frac{\tau_L}{\tau_C}}$ $\le {\cal {E}}_0$. As a result, the chirp 
does not change   the energy $E_{pulse}$ carried by the field, which is  proportional to the square of the amplitude ${\cal{E}}_0$ and to the temporal width $\tau_L$ of the transform limited pulse: $E_{pulse}$=$\frac{c \epsilon_0}{2}\int_{-\infty}^{+\infty} |{\cal {E}}(t)|^2 dt$= $ \frac{I_L\tau_L }{2} \sqrt\frac{\pi}{\ln 2}$, with $I_L=\frac{c \epsilon_0}{2} {\cal {E}}_0^2$.

The instantaneous  frequency $\omega(t)/2\pi$ is given by the derivative of the rapidly oscillating term in ${\cal {E}}(t)$: 
\begin{equation}
\omega(t)=\omega_L +\frac{d\varphi}{dt}=\omega_L +\chi (t-t_P),
\label{eq:freq}
\end{equation}
and varies linearly around the central frequency $\omega_L/2\pi$.

\subsection{The two-channel coupled equations}
We study the dynamics in the ground and excited states coupled by the pulse described in Eq.(\ref{eq:field}), for which we consider linear polarization $\vec{e_L}$. In the dipole approximation, the coupling term between the two electronic channels is written as:
$ -\vec{D}_{ge}({R}) \cdot \vec{e_L} {\cal {E}}(t)$ $\approx D_{ge}^{\vec{e_L}} {\cal {E}}(t)$,
where  $\vec{D_{ge}}(R)$  is the R-dependent matrix element of the dipole moment operator  between the ground and the excited molecular electronic states. Since  the photoassociation reaction occurs at large distances ($R \ge 90$ a$_0$),  we  neglect the $R-$dependence,  using the asymptotic value  $D_{ge}^{\vec{e_L}}$ deduced from standard long-range calculations \cite{vatasescu99}.

In the rotating wave approximation with the instantaneous frequency, the coupled equations for the radial wavefunctions $\Psi^{\omega}_{g,e}(R,t)$ in the ground and excited states can be written as \cite{eluc04}:
\begin{eqnarray} 
\label{eq:cplfin}
&&i\hbar\frac{\partial}{\partial t}\left(\begin{array}{c}
 \Psi^{\omega}_{e}(R,t)\\
\Psi^{\omega}_{g}(R,t)
 \end{array}\right)=\\
&&
\left(\begin{array}{lc}
 {\bf \hat T} + \bar{V}(R)+ \Delta (R,t)  & 
W_L f(t) \\
 W_L f(t) & 
 {\bf \hat T} + \bar{V}(R)-\Delta (R,t)
 \end{array} \right) 
 \left( \begin{array}{c}
 \Psi^{\omega}_{e}(R,t)\\
\Psi^{\omega}_{g}(R,t) 
 \end{array} \right) \nonumber
 \end{eqnarray}
Eq. (\ref{eq:cplfin}) corresponds to a rotating-frame transformation, with the angle $\omega(t)$, leading  to a ``frequency-modulated frame'' \cite{goswami02}. The coupling between the two channels writes as:
\begin{eqnarray}
W_Lf(t) \le W_L \sqrt{\frac{\tau_L}{\tau_C}}=W_{max}, \\
W_L= - \frac {1}{2}{\cal {E}}_0 D_{ge}^{\vec{e_L}}= - \frac {1}{2} \sqrt{\frac{2I}{c \epsilon_0}} 
D_{ge}^{\vec{e_L}}
\label{eq:couplt}
\end{eqnarray}
The diagonal terms of the hamiltonian matrix contain the kinetic energy operator ${\bf \hat T}$, the mean potential $\bar{V}(R)$:
\begin{equation}
\bar{V}(R)=\frac{V_{exc}(R)+V_{ground}(R)}{2},
\end{equation}
and the R- and t-dependent energy difference $\Delta(R,t)$ between the potentials dressed by the instantaneous laser energy $\hbar\omega(t)$ (defined in Eq. \ref{eq:freq}):
\begin{equation}
2\Delta (R,t)=2\Delta_L(R)- \hbar\frac{d \varphi}{dt}=2\Delta_L(R)-\hbar \chi(t-t_P), 
\label{eq:Del}
\end{equation}
where $2\Delta_L(R)$ is the R-dependent energy difference between the two electronic potentials dressed by the mean laser energy $\hbar\omega_{L}$ and which are crossing in $R_L$:
\begin{eqnarray}
2 \Delta_L(R)= V_{exc}(R)- V_{ground}(R)-\hbar\omega_{L},\\
2 \Delta_L(R_L)=0, \ 2 \Delta_L(R \to \infty) \to \delta_L^{at}
\label{eq:Delta}
\end{eqnarray}
 The instantaneous crossing point $R_C(t)$ between the two electronic dressed potentials coupled by the pulse is defined by:
\begin{equation}
 \Delta (R_C(t),t)=0 ; \ \ R_C(t_P)=R_L
\label{eq:reschirp}
\end{equation}

\subsection{Choice of the pulse parameters}

The choice of the present paper is to fix the carrier frequency  $\omega_L/2 \pi$ (or, equivalently, the detuning $\delta^{at}_{L}$), the coupling $W_L$ (or the intensity $I_L$), and the chirp rate $\chi$ in the time domain. In exchange, we shall study pulses of different durations $\tau_L$ (or various spectral widths $\delta \omega$) leading to different swept energy ranges 2$\hbar |\chi| \tau_C$,  maxima $W_{max}=W_L \sqrt{\frac{\tau_L}{\tau_C}}$ of the coupling, and energies $E_{max} \sim W_L^2 \tau_L $.

Fixing $W_L$, instead of $E_{max}$, leads to situations as various as possible for either the photoassociation yield or the dynamics of the process. Indeed, the pulses which are studied induce quite different excitation regimes, due to their characteristics: narrow or broad bandwidth $\delta \omega$, short or long duration $\tau_C$, small or large chirp $\Phi^{\prime\prime}$ inducing various ratios $\frac{\tau_L}{\tau_C}$ and various energies $E_{max}$. The object of this work is to analyze the properties of the photoassociated wavepacket as  a function of the pulse. In this aim we shall study the  photoassociation yield and the radial distribution of population transferred in the $0_g^-$ surface at the end of the pulse, but also the evolution of the wavepackets after the pulse (vibration in the 0$_g^-$ potential and acceleration to the inner region). An important goal is to gain information about the degree of coherent control in the photoassociation process: for example we are interested to maximize the population localized at  small distances in the $0_g^-$ potential (internuclear distances more favorable to radiative stabilization of the photoassociated molecules, by spontaneous or stimulated emission, towards the ground potential surface).

$\bullet$ \underline{\it Detuning.}
 We present here calculations for photoassociation with different pulses at the same detuning  $\delta_L^{at}$=2.656 cm$^{-1}$, corresponding  to a crossing point $R_L=93.7 \ a_0$. The detuning is chosen such as $\delta_L^{at}=E_{v_0}$, the binding energy of the $v_0$=98 level in the external well of the 0$_g^-(6s+6p_{3/2})$ potential, which means that  at $t=t_P$ there is a resonance condition between the continuum state lying at threshold (E=0) and this vibrational level. Taking into account that the initial collisional state considered here corresponds to a very low temperature T=54 $\mu$K, and to an energy $E_0=k_BT$ much smaller than the vibrational energy spacing in the 0$_g^-(6s+6p_{3/2})$ potential ($\frac{E_{v_0} -E_{v_0-1}}{E_0} \approx 3300$), the initial state $E_0$  is in resonance with the $v_0$ vibrational level.

$\bullet$ \underline{\it Intensity $I_L$ and Coupling $W_L$.}
Also, we consider the same laser intensity $I_L$ and then the same coupling $W_L$ for all the pulses. But the pulses with different durations $\tau_L$ correspond to different pulse energies  $E_{pulse} \sim I_L\tau_L $. For laser excitation with $\pi$ polarization between the electronic states $^3\Sigma_u^+(6s,6s)$ and 0$_g^-(6s+6p_{3/2})$, and neglecting the $R$-variation of the dipole coupling, the intensity is related to the coupling $W_L$ by Eq. (\ref{eq:couplt}), giving $W_L$(a.u.)=$9.74 \times 10^{-9}\sqrt{I_L(W/cm^2)}$ \cite{vatasescu99}. In the present calculations the laser intensity is $I_L=120$ kW cm$^{-2}$, giving a  coupling $W_L$=0.7396 cm$^{-1}$. By chirping the pulse, the instantaneous coupling becomes $W(t)=W_L f(t) \le W_{max}$.

$\bullet$ \underline{\it Chirp rate $\chi$ in the time domain. Focussing.}
The linear chirp parameter  $\chi$  has been designed in order to achieve, at a time $t=t_P + T_{vib}(v_0)/2$ (where $T_{vib}(v_0)/2$= 125 ps is half the vibrational period of the level $v_0$=98), the focussing of the excited vibrational wavepacket at the internal turning point of the vibrational state $v_0$ in the $0_g^-$ outer well.  The chirp parameter necessary to compensate the dispersion in the vibrational period of the wavepacket is chosen as $\chi=- 2 \pi T_{rev}(v_0)/[T_{vib}(v_0)]^3$, i.e. adjusted to match the revival period $T_{rev}$ (see Appendix B) of the resonant level $v_0$=98.  This leads to $\chi$=-4.79$\times$ 10$^{-3}$ ps$^{-2}$=-0.28 $\times$ 10$^{-11}$ au, and $\hbar\chi$=-0.025 cm$^{-1}$ ps$^{-1}$.

$\bullet$ \underline{\it Chirp rate $\Phi^{\prime\prime}$ in the frequency domain.}
By combining the relation (\ref{eq:chirparam}) with Eq. (\ref{eq:stretch}) one obtains a second degree equation for $\Phi^{\prime\prime}$:
\begin{equation}
\Phi^{\prime\prime 2} -\frac{\Phi^{\prime\prime}}{\chi}+ \frac{\tau_L^4}{(4 \ln 2)^2}=0
\label{eq:phi2eq}
\end{equation}
This means that for an initial pulse of temporal width $\tau_L$, there are two values of the chirp $\Phi^{\prime\prime}$ in the frequency domain corresponding to the same chirp  $\chi$ in the time domain:
\begin{equation}
\Phi^{\prime\prime}_{1,2}= \frac{1}{2|\chi|} \left( \frac{|\chi|}{\chi}\pm \sqrt{1-\frac{\chi^2 \tau_L^4}{(2 \ln 2)^2}} \right),
\label{eq:phi12}
\end{equation}
which  give two different temporal widths $\tau_C$, as it can be seen with the formula (\ref{eq:stretch}).
$\bullet$ \underline{\it Pulse duration $\tau_C$.}
 $\tau_L$ and  $\chi$ being chosen,  $\tau_C^2$ is solution of the second order equation:
\begin{equation}
\tau_C^4 -\frac{(4 \ln 2)^2}{\chi^2 \tau_L^2}\tau_C^2 + \frac{(4 \ln 2)^2}{\chi^2} =0
\label{eq:tauc2eq}
\end{equation}
which  gives, as for $\Phi^{\prime\prime}$, two values of the pulse duration $\tau_C$ for a given pair  ($\tau_L$, $\chi$). In fact, with the relation (\ref{eq:chirparam}) one can see that the pulse with the longer duration  $\tau_C$ corresponds to the higher rate  $\Phi^{\prime\prime}$ in the frequency domain. 

$\bullet$ \underline{\it Temporal width $\tau_L$ of the pulse before chirping.}
The Eqs. (\ref{eq:phi2eq},\ref{eq:tauc2eq}) show that the existence of real values of $\Phi^{\prime\prime}$ and  $\tau_C$ requires:
\begin{equation}
\tau_L \le  \sqrt{\frac{2 \ln 2}{|\chi|}},
\label{eq:condtaul1}
\end{equation}
imposing an {\it upper limit} for the value of $\tau_L$ which can still be chosen for a fixed value of  $\chi$. 
On the other hand, one can choose to avoid the excitation of the $0_g^-$ continuum during the photoassociation process. Such a requirement is related to a {\it lower limit} for the initial temporal width $\tau_L$, coming from a  condition for a spectral width of the pulse  smaller than the detuning $\delta^{at}_{L}$ corresponding to the vibrational level resonant at $t=t_P$:
\begin{equation}
\hbar \delta \omega =\hbar\frac{4\ln 2}{\tau_L} < \delta^{at}_{L} \\
 \Longrightarrow \tau_L > \hbar\frac{4\ln 2}{\delta^{at}_{L}}
\label{eq:condtaul2}
\end{equation}
For the detuning $\delta_L^{at}$=2.656 cm$^{-1}$ and the chirp rate $\chi$=-4.79$\times$ 10$^{-3}$ ps$^{-2}$ leading to focussing, the conditions 
(\ref{eq:condtaul1}) and (\ref{eq:condtaul2}) give the following interval of choice for $\tau_L$:
\begin{equation}
5.54 \ ps < \tau_L \le 17 \ ps,
\label{eq:intervtaul}
\end{equation}
corresponding to predominant excitation of bound vibrational levels  $0_g^-$.

For $0 < \tau_L <  5.54 \ ps$, dissociation continuum states of $0_g^-$ localized at large distances are mainly excited. Let us remark that the frontier between excitation of bound or continuum states is not strict, it only corresponds to the FWHM of the spectral energy distribution, without taking into account the absolute intensity of the pulse. In the interval described by the relation (\ref{eq:intervtaul}), the excitation of continuum will become important for an increasing intensity $I_L$. The characteristics of the pulses studied in the present paper are reported in the Table 1, which also shows those of the pulse studied in Ref. \cite{eluc04}.

\begin{figure*}
\resizebox{0.75\textwidth}{!}{%
\includegraphics{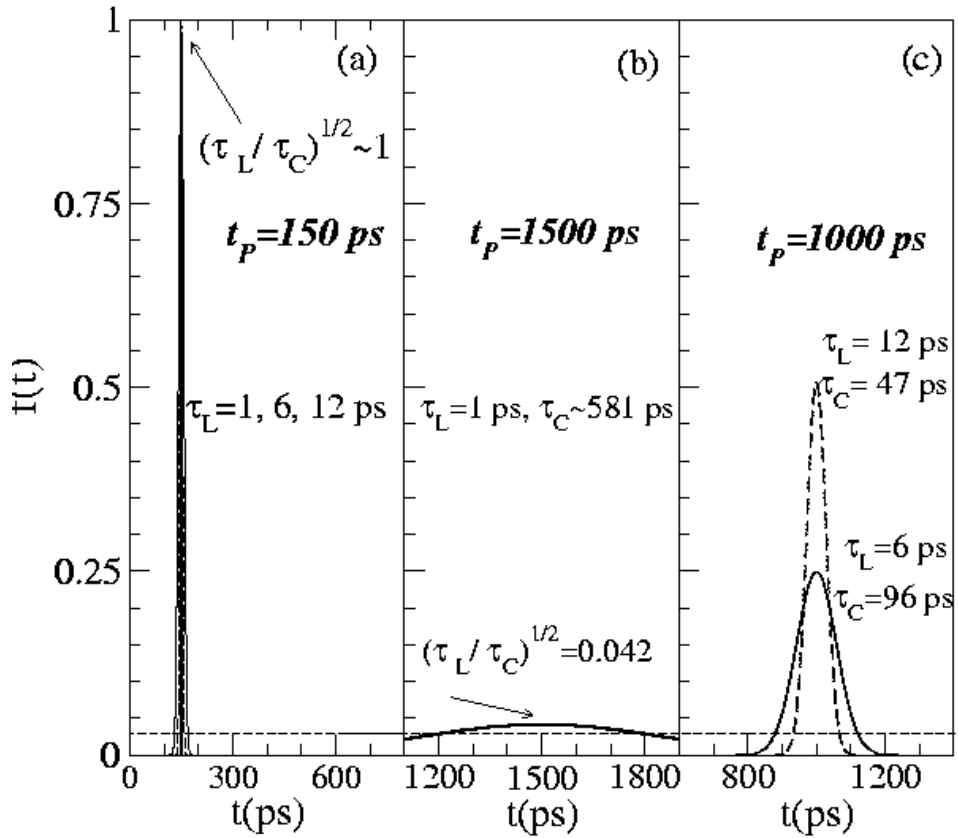}
}
\caption{Time-dependent gaussian envelopes f(t) of the linearly chirped pulses considered in this paper, all having the same chirp rate $\chi=-4.79 \times 10^{-3} ps^{-2}$ in the time domain, but different $\tau_L$ values, stretched to $\tau_C$. At each chosen $\tau_L$ value correspond two values $\tau_C$, as explained in the text. The maximum of the pulse, $f(t_P)=\sqrt{\frac{\tau_L}{\tau_C}}$, decreases when the time width $\tau_C$ increases. a) Chirped pulses with $\tau_C \approx \tau_L$= 1 ps, 6 ps and 12 ps, centered at $t_P$=150 ps. b) Chirped pulse with $\tau_C$=581 ps ($\tau_L$= 1 ps), centered at $t_P$=1500 ps. c) Chirped pulses with $\tau_C$=96 ps ($\tau_L$= 6 ps)  and $\tau_C$=47 ps ($\tau_L$= 12 ps), both centered at $t_P$=1000 ps. The horizontal broken line indicates the value from which the pulse becomes efficient in the population transfer.
 For the same intensity $I_L$ of the laser, the pulses having the same initial $\tau_L$ correspond to the same energy $E_{pulse}$= $ \frac{I_L\tau_L }{2} \sqrt\frac{\pi}{\ln 2}$, carried by the field.}
\label{fig:pulses-env}
\end{figure*}

\subsection{The energy range corresponding to a large population transfer induced by the chirped pulse}
\label{sec:enrange}
The impulsive approximation \cite{banin94,eluc04} can be used to qualitatively  determine the spatial range in which the population transfer from the ground to the excited surface is important. Let's take the example of a two-level system with energy splitting $\hbar \omega_{eg}= \hbar( \omega_{e}- \omega_{g})$ which is excited by a pulse characterized by a spectral energy distribution $|\tilde{\cal E}(\omega)|^2$ (centered on the mean laser  pulsation  $\omega_{L}$, FWHM $\delta \omega$, and maximum proportional to the laser intensity I). Longtime after the end of the pulse, and in the perturbative approximation, the population in the excited level is given by $\mu_{ge}^2 |\tilde{\cal E}(\omega_ {eg})|^2$, where $\mu_{ge}$ is the dipole moment of the transition. Therefore this  population grows up when $\mu_{ge}$ or I increase and when the detuning $\omega_{eg}- \omega_{L}$ decreases, being maximum at the resonance ( $\omega_{eg}= \omega_{L}$). 

Similar considerations apply for population transfer from the ground to the excited surface; it is favoured: (i) at resonance (i.e. at the instantaneous crossing point $R_c(t)$), (ii) for a large overlap integral $|<^3\Sigma_u^+ E_0| 0_g^- v>|^2$ between the initial collisional wavefunction and the resonantly excited vibrational level $v$ having its outer turning point close to $R_c(t)$, and (iii) with the increase of the laser intensity.

For a gaussian pulse, 98 $\%$ of the energy is carried during the time interval $[-\tau_C, +\tau_C]$, and then the effect of the pulse can be estimated  by analyzing it during this {\it temporal window} \cite{eluc04}. During this time interval, the instantaneous crossing point  $R_C(t)$ defined by the relations (\ref{eq:reschirp}) varies in time, describing a {\it resonance window}; the energy range resonantly swept  by the pulse around the central frequency $\omega_L/2 \pi$ is 2$\hbar |\chi| \tau_C$, being limited by its spectral width \cite{eluc04}. Indeed, $ |\chi| \tau_C \le \delta \omega$, with a ratio depending on the chirp rates:
\begin{equation}
\frac {|\chi| \tau_C}{\delta \omega}=  \sqrt{\Phi^{\prime\prime}\chi}=
\sqrt{1-\left(\frac{\tau_L}{\tau_C} \right)^2 } \le 1
\label{eq:tauchi}
\end{equation}
Then, vibrational levels lying in the ``resonance window'' will be excited by the pulse at different moments.

Besides, the large overlap of the initial continuum wavefunction with  the  $0_g^-$ vibrational wavefunctions of levels close to the dissociation limit is especially favorable to population transfer at  very large distances. Indeed, the overlap integral has a large maximum for vibrational levels having outer turning points at large distances R, close to the dissociation limit (see Appendix C, Fig. \ref{fig:recouvr}). 

Furthermore, for a large laser intensity, off-resonant excitation can easily become efficient. This is particularly true for cold photoassociation at small detunings if the coupling and detuning have comparable values, leading to noticeable excitation outside the resonance window. 

 Then, it is interesting to note that using a chirped pulse for such free-bound transitions close to the dissociation limit, it is possible to control the excited energy range  by the pulse characteristics, in order to avoid the population of the continuum or to restrain the spatial range covered by the final $0_g^-$ packet. Indeed, in addition to the spectral width $\delta \omega$ (determined by $\tau_L$ for both transform limited or chirped pulses), one can select the range of the window $|\chi| \tau_C$ to be excited resonantly, by choosing the chirp parameters $\Phi^{\prime\prime}$, $\chi$, or, equivalently, the ratio $\frac{\tau_L}{\tau_C}$ (see Eq. (\ref{eq:tauchi})).

Lastly, the pulse intensity $I_L$ will determine the adiabatic or non-adiabatic character of the transfer (this  will be detailed in Sec. \ref{sec:adiabwind}), and the importance of the off-resonance excitation.  Nevertheless, the estimated boundary between resonant and off-resonant excitation remains approximative. 

\begin{table*}
\label{tab:pulses}
\caption{Parameters of gaussian pulses considered in this work, linearly chirped with the same rate $\chi$ in the time domain, and corresponding to the laser intensity at $t=t_P$ for the transform limited pulse $I_L=120$ kW cm$^{-2}$, which, for a linear polarization of the electric field, gives a coupling $W_L$=0.7396 cm$^{-1}$ for the $a^3\Sigma_u^+(6s,6s)$ $\to$ $0_g^-(6s,6p_{3/2})$ transition at large interatomic distances. The central frequency of the pulse, $\omega_L/2 \pi$, resonantly excites the $v_0=98$ level of the $0_g^-(6s,6p_{3/2})$ outer well, with binding energy $E_{v_0}=2.656$ cm$^{-1}$ and  vibrational period $T_{vib}(v_0)=250$ ps. The significant parameters listed below are: temporal width $\tau_L$ of the initial pulse before chirping, spectral width $\hbar\delta \omega$, chirp rate $\Phi^{\prime\prime}$ in the frequency domain, temporal width $\tau_C$ of the chirped pulse, the ratio $\sqrt{\frac{\tau_L}{\tau_C}}$, the chirp rate $\chi$ in the time domain, the energy range $\hbar|\chi|\tau_C$ resonantly swept by the pulse during the period $[-\tau_C, \tau_C]$, the maximum coupling
 $W_{max}=W_L\sqrt{\frac{\tau_L}{\tau_C}}=W(t_P)$, and the parameter $\alpha_{max}$ indicating a limit for the adiabaticity range $[-\alpha \tau_C, \alpha \tau_C]$ ($\alpha \ll\alpha_{max}$). The probability of photoassociation in the $0_g^-$ surface at the end of the each  pulse (corresponding to a total population normalized at 1 on the grid), noted $P_{0_g^-}(E_0)$, is also shown ($E_0$ is the energy of the initial continuum state, $E_0/k_B=54$ $\mu K$). The probability per pump pulse that a given pair of atoms at the temperature $T=E_0/k_B$ to be photoassociated is roughly equal to ${\cal P}_{0_g^-}(T)$$=P_{0_g^-}(E_0) \frac{k_B T}{(\partial E/\partial n) |_{E_0}} \frac{1}{Z}$ $\approx 20 P_{0_g^-}(E_0)\frac{1}{Z}$, where $k_B T$ is the width of the thermal distribution and $(\partial n/\partial E) |_{E_0}$ the density of collisional states at the energy $E_0$ ($Z$ being the partition function, see Eq.(\ref{eq:partitionz})) . The characteristics of the chirped pulse studied in Ref. \cite{eluc04} are reported in the last line.} 
\begin{tabular}{|l|c|lll|l|c|c|c|c|}
\hline\noalign{\smallskip}
&$\hbar \delta \omega$ (cm$^{-1}$)&$\Phi^{\prime\prime}$ (ps$^2$)&{\bf $\tau_C$} (ps)&$\sqrt{\frac{\tau_L}{\tau_C}}$&$\chi$ (ps$^{-2}$)&$\hbar|\chi|\tau_C$ (cm$^{-1}$)&$W_{max}$ (cm$^{-1}$)&$\alpha_{max}$&$P_{0_g^-}(E_0)$\\ 
\noalign{\smallskip}\hline\noalign{\smallskip}
 $\tau_L$= {\bf 1 ps}&14.72 &-0.00063&{\bf 1.002}&0.999&-4.79$\times$ 10$^{-3}$&0.025&0.739&1.12&3.961 $\times$ 10$^{-2}$\\
&&-208.8&{\bf 581}&0.042&&14.52&0.031&-&3.941 $\times$ 10$^{-2}$ \\
\hline\noalign{\smallskip}
$\tau_L$= {\bf 6 ps}&2.453&-0.81 &{\bf 6.012}&0.999&-4.79$\times$ 10$^{-3}$&0.15&0.739&1.12&2.168 $\times$ 10$^{-2}$\\
&&-207.99&{\bf 96.28}&0.249&&2.41&0.184&0.5&6.005 $\times$ 10$^{-2}$\\
\hline\noalign{\smallskip}
$\tau_L$= {\bf 12 ps}&1.227&-13.84&{\bf 12.41}&0.983&-4.79$\times$ 10$^{-3}$&0.31&0.727&1.11&3.046 $\times$ 10$^{-4}$\\
&&-194.96 &{\bf 46.6}&0.507&&1.17&0.375&0.87&5.340 $\times$ 10$^{-4}$\\
\hline\noalign{\smallskip}
$\tau_L$= {\bf 15 ps}&0.981&-38.51&{\bf 16.60}&0.950&-4.79$\times$ 10$^{-3}$&0.42&0.703&1.10&1.200 $\times$ 10$^{-4}$\\
&&-170.00&{\bf 34.8}&0.657&&0.87&0.486&0.95&3.245 $\times$ 10$^{-4}$\\
\noalign{\smallskip}\hline
\end{tabular}
\vspace*{1cm}
\end{table*}

\subsection{Condition for an ``adiabaticity window'' during the pulse duration}
\label{sec:adiabwind}

Efficient adiabatic population inversion can be obtained by using a chirped laser pulse, which in a two state system allows to sweep the instantaneous  frequency $\omega(t)$ from far above (respectively far below) to far below (respectively far above) resonance. Sufficiently slow sweeping induces total adiabatic transfer from one state to the other one \cite{goswami02,cao98,cao00,eluc04}. 

In a previous work \cite{eluc04} we have analyzed adiabatic population inversion within the impulsive limit \cite{banin94}, assuming that the relative motion of the two nuclei is frozen during the laser interaction, i.e. $\tau_C \ll T_{vib}(v_0)$, where $T_{vib}(v_0)=250$ ps is the vibrational period of the $0_g^-$ level resonantly populated at $t=t_P$. By neglecting the kinetic energy operator appearing in the two-level Hamiltonian of the Eq. (\ref{eq:cplfin}), one can  introduce  a coordinate-dependent two-level model able to define the conditions for full adiabatic population transfer. At the instantaneous crossing points  $R_C(t)$, the nonadiabatic effects can be explored using the  adiabaticity condition, which in these points takes a simple form determined by the pulse shape ($|\chi|$ and $\frac{\tau_C}{\tau_L}$) and its intensity ($I_{L}$ or $W_{L}$) \cite{eluc04}:
\begin{equation}
\hbar^2 |\chi| \ll 8(W(t))^2.
\label{adiabcross}
\end{equation}
The condition (\ref{adiabcross}) will not  be verified  when W(t) becomes very small.  Then, it is useful to estimate the domain $[-\alpha \tau_C, \alpha \tau_C]$ , with  $\alpha>0$, for which the adiabaticity condition (\ref{adiabcross}) is verified. For $|t-t_P|< \alpha\tau_C$, the coupling  W(t) has the lower bound $\frac{1}{4^{\alpha^2}} W_{max}$, so $\alpha$ can be deduced from the condition:
\begin{equation}
\frac{16^{\alpha^2}}{8} \hbar^2 |\chi| \frac{\tau_C}{\tau_L} \ll  W_{L}^2,
\label{eq:adiabcond3}
\end{equation}
giving:
\begin{equation}
\alpha \ll \sqrt{\frac{1}{\ln 16} \ln \lbrace 8 \frac{1}{\hbar^2 |\chi|} W_{L}^2 \frac{\tau_L}{\tau_C}} \rbrace=\alpha_{max}
\label{eq:alpha}
\end{equation}
For $\alpha \ll \alpha_{max}$, the adiabaticity condition (\ref{eq:adiabcond3}) is very well satisfied during all the time interval $[-\alpha \tau_C, \alpha \tau_C]$. For $\alpha \approx \alpha _{max}$, the transfer can be adiabatic during a certain portion of this interval, but strong non-adiabatic effects appear at the boundary $|t-t_P| \approx \alpha_{max} \tau_C$.
For $8 \frac{1}{\hbar^2 |\chi|} W_{L}^2 \frac{\tau_L}{\tau_C} <1$ (small intensity $I_{L}$ and/or large chirp $|\chi| \frac{\tau_C}{\tau_L}$), adiabaticity never occurs, $\alpha_{max}$ not being defined. The values $\alpha_{max}$ corresponding to the  pulses studied here are given in the Table \ref{tab:pulses}. 
  
For a given pulse,  $\alpha_{max}$ offers an approximate evaluation for the extension of the time interval  $[-\alpha \tau_C, \alpha \tau_C]$, with $\alpha \ll \alpha_{max}$, during which the adiabaticity condition can be verified at the instantaneous crossing points. Then, for fixed $|\chi|$ and $W_{L}$, it is the ratio $\frac{\tau_L}{\tau_C}$ which fixes the time-interval during which crossing of the two potentials  can lead to population inversion due to adiabatic passage  \cite{cao00,goswami02,eluc04}.

\section{Results for pulses having different durations}
 We consider pulses generated with three initial temporal widths: $\tau_L$= 1 ps, 6 ps and 12 ps, and which are chirped with the same chirp rate in the time domain: $\chi$=-4.79$\times$ 10$^{-3}$ ps$^{-2}$. The characteristics of the pulses are given in the Table 1. The impulsive limit is valid for all these pulses, except the pulse with $\tau_C$=581 ps ($\tau_L$= 1 ps).

As discussed before, to each given value $\tau_L$ (determining the spectral width
 $\delta \omega \sim 1/\tau_L $ and the total energy of the pulse $E_{pulse} \sim I_L \tau_L$) correspond two chirped pulses with very different characteristics: one has $\tau_C \approx \tau_L$, being ``almost no-chirped'',  and the other is a ``really chirped'' one, having a much larger temporal width than the initial pulse. The envelopes of the 6 pulses thus built  are shown in Fig. \ref{fig:pulses-env}. For each pulse we shall analyze the resulting photoassociation dynamics: the relative population yield transferred to the $0_g^-$ state (for a total population normalized at 1 on the grid), its radial distribution, and the modification of the density probability  in the initial $E_0/k_B=54$ $\mu K$ $a^3\Sigma_u^+(6s,6s)$  state.
 Our aim is to analyze their efficiency for the photoassociation reaction, the main criterion being a maximum transfer of population from the initial $a^3\Sigma_u^+(6s,6s)$ $E_0$ continuum state to bound vibrational levels of the $0_g^-(6s,6p_{3/2})$ excited state: this means that we are interested in those pulses able to bring the population located at large distances in the initial continuum and to put it in the $0_g^-(6s,6p_{3/2})$ external well, in such a manner that the most of this population to be present at small distances (in particular at the inner turning point of the outer well) with a small delay after the end of the pulse, or to be efficiently accelerated to the inner region (in the case of the population transferred at large distances). Maximizing the population transferred to the inner region of the excited potential ($R<100$ a$_0$) is a first step in a process leading to an efficient formation of cold molecules. 

The comparison between the pulses  described before can be made from two points of view: i) one can compare ``really- chirped'' and ``almost no-chirped'' pulses, by looking at results given by the pulses with the same $\tau_L$ (the same spectral width $\hbar \delta \omega$ and the same energy $E_{pulse}$), and ii) one can analyze the main differences between the results given by short pulses with a large spectral bandwidth and those given by much longer pulses with a narrower spectral width.

Several comments are to be made about what we could expect from these pulses, given the values illustrated in the Table \ref{tab:pulses}, and the fact that the detuning corresponding to the $0_g^-(6s,6p_{3/2})$ level $v_0=98$  excited at $t=t_P$ is $\delta_L^{at}$=2.656 cm$^{-1}$.\\ 
First, we can note that the shorter pulses with $\tau_L$=1 ps and $\tau_L$=6 ps  have the spectral widths  $\hbar \delta \omega$= 14.72 cm$^{-1}$, much larger than $\delta_L^{at}$=2.656 cm$^{-1}$, and 2.453 cm$^{-1}$, of the same order of magnitude as $\delta_L^{at}$. This means that, for $\tau_L$=1 ps,  the $0_g^-(6s,6p_{3/2})$ continuum will be massively excited. The pulses with $\tau_L$= 12 ps have the spectral bandwidth $\hbar \delta \omega$= 1.227 cm$^{-1}$ smaller than $\delta_L^{at}$, which avoids to noticeably populate the $0_g^-$ continuum (except if the intensity is too strong, see Sec. \ref{sec:enrange}).\\ 
Second, the horizontal line in Fig. \ref{fig:pulses-env} indicates the value from which the instantaneous coupling strength $W_L f(t)$ becomes efficient in the population transfer. The ``efficient time durations'' of the studied pulses are generally of the same order of magnitude as their corresponding $\tau_C$, excepting the pulse with $\tau_C=581$ ps ($\tau_L$=1 ps), for which its active interval  is reduced by a factor of $\approx 4$: indeed, due to the strong diminuation of $W_{max}$ by the chirp (small ratio $\sqrt{\frac{\tau_L}{\tau_C}}$) only a small fraction of the pulse represented in Fig. \ref{fig:pulses-env}b) can produce population transfer.\\
Another aspect refers to the value of the detuning $\delta_L^{at}/2=1.328$ cm$^{-1}$ which is always larger than the  maximum coupling $W_{max}$ of the studied pulses (see the Table 1). Therefore at  $t=t_P$, the off-resonant Rabi coupling at very large distances $R_{\infty}$ \cite{eluc04}:
\begin{equation}
\hbar \Omega(t_P,R_{\infty})=\sqrt{W^2_{max}+ (\frac{\delta_L^{at}}{2})^2 }
\label{eq:Rabicoupl}
\end{equation}
will be larger than the resonant coupling $W_{max}$ at $R_L$ (indeed, for the pulses listed in the Table 1, the values of $\hbar \Omega(t_P,R_{\infty})$ are between 1.3 and 1.4 cm$^{-1}$, the corresponding Rabi periods $T_{Rabi}=\pi / \Omega$ varying between 11 and 12.5 ps). This means that, for this small detuning, the field is strong enough to couple the two channels at very large distances, well beyond we could expect from the estimated energy range $2\hbar|\chi|\tau_C$ resonantly swept by the pulse during the period $[-\tau_C, \tau_C]$. In fact, the off-resonance excitation at very large distances is enforced by several factors: the small detuning $\delta_L^{at}$ (corresponding to $t=t_P$), the coupling $W_L$, but also the much bigger overlap of the vibrational wavefunctions $0_g^-$ $v$ with the initial continuum $a^3\Sigma_u^+$ $E_0$ (see Appendix C, Fig.\ref{fig:recouvr}), which for $v=160 \to 170$ ($E_v=-0.4 \to -0.2$ cm$^{-1}$) is about 3.7 times larger than the overlap with the  state  $v_0=98$ ($E_{v_0}=-2.6 $ cm$^{-1}$). Lastly, the negative chirp presently studied  begins resonant excitation at $t<t_P$ from the higher vibrational levels $v>98$ of the $0_g^-(6s,6p_{3/2})$ state. These weakly bound vibrational levels,  which are excited before the maximum of the pulse at $t=t_P$,  are much more sensitive at the presence of the field than the levels excited at  $t>t_P$. Indeed, the time evolution of the population excited in these states shows Rabi oscillations, which are the signature of significant nonadiabatic effects, as discussed in Ref. \cite{eluc04}.

\subsection{Population transferred to the $0_g^-(6s,6p_{3/2})$ excited state}
We shall analyze  the population transferred in the $0_g^-(6s,6p_{3/2})$ state during the photoassociation process, using the pulses whose envelopes f(t) are represented in Fig. \ref{fig:pulses-env}. The results shown in this section correspond to a total population (initially  in the $a^3\Sigma_u^+(6s,6s)$  state) which  is normalized at 1 on the whole grid of extension $L_R$. We represent the following quantities:

$\bullet$ the time evolution of the probability for population transfer on the whole grid in the $0_g^-(6s,6p_{3/2})$ surface:
\begin{equation}
P_{0_g^-}(t) = \int_{0}^{L_R} | \Psi_{0_g^-}(R',t) |^{2} dR'
\label{eq:p0gt}
\end{equation}
The relative population $P_{0_g^-}(E_0)$ transferred in the $0_g^-(6s,6p_{3/2})$ channel after the end  of the pulse ($t-t_P \gg \tau_C$) is given in the Table 1. The probability of photoassociation per pump pulse of a pair of atoms at the temperature $T=E_0/k_B$ is roughly equal to ${\cal P}_{0_g^-}(T)$$=P_{0_g^-}(E_0) \frac{k_B T}{(\partial E/\partial n) |_{E_0}} \frac{1}{Z}$ $\approx 20 P_{0_g^-}(E_0)\frac{1}{Z}$, where $k_B T$ is the width of the thermal distribution and $(\partial n/\partial E) |_{E_0}$ the density of collisional states at the energy $E_0$ ($Z$ being the partition function, see Eq.(\ref{eq:partitionz})) 

$\bullet$ the distribution of the  $0_g^-(6s,6p_{3/2})$ radialy integrated population, $P_{0_g^-}(R,t_{foc})$,  as a function of the distance R, at the moment 
$t_{foc}=t_P+\frac{T_{vib}}{2}=t_P$+130 ps, corresponding to the focussing at the inner turning point. At this time moment almost all the studied pulses  are practically ``finished'', $W(t) \approx 0$, except the very large pulse with $\tau_C$= 581 ps ($\tau_L$= 1 ps), 
for which anyway the efficient time leading to significant transfer is over (see the dashed line in Fig. \ref{fig:pulses-env}). For $t>t_{foc}$ the population $P_{0_g^-}$ remains constant, the two potentials $0_g^-$ and $a^3\Sigma_u^+$ being no longer coupled.
\begin{equation}
P_{0_g^-}(R,t_{foc}) = \int_{0}^{R} | \Psi_{0_g^-}(R',t_{foc}) |^{2} dR'
\label{eq:p0gR}
\end{equation}

\begin{figure}
\resizebox{0.5\textwidth}{!}{%
\includegraphics{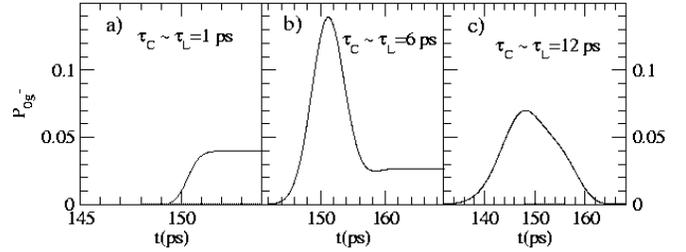}
}
\caption{Time evolution of the $0_g^-(6s,6p_{3/2})$ population (for a total population normalized at 1 on the grid)  obtained by photoassociating with the ``almost no-chirped'' pulses having $\tau_C \approx \tau_L$, whose envelopes f(t), represented in Fig. \ref{fig:pulses-env}a), are centered at $t_P=150$ ps.}
\label{fig:pop0g-nc3}
\end{figure}
\begin{figure}
\resizebox{0.5\textwidth}{!}{%
\includegraphics{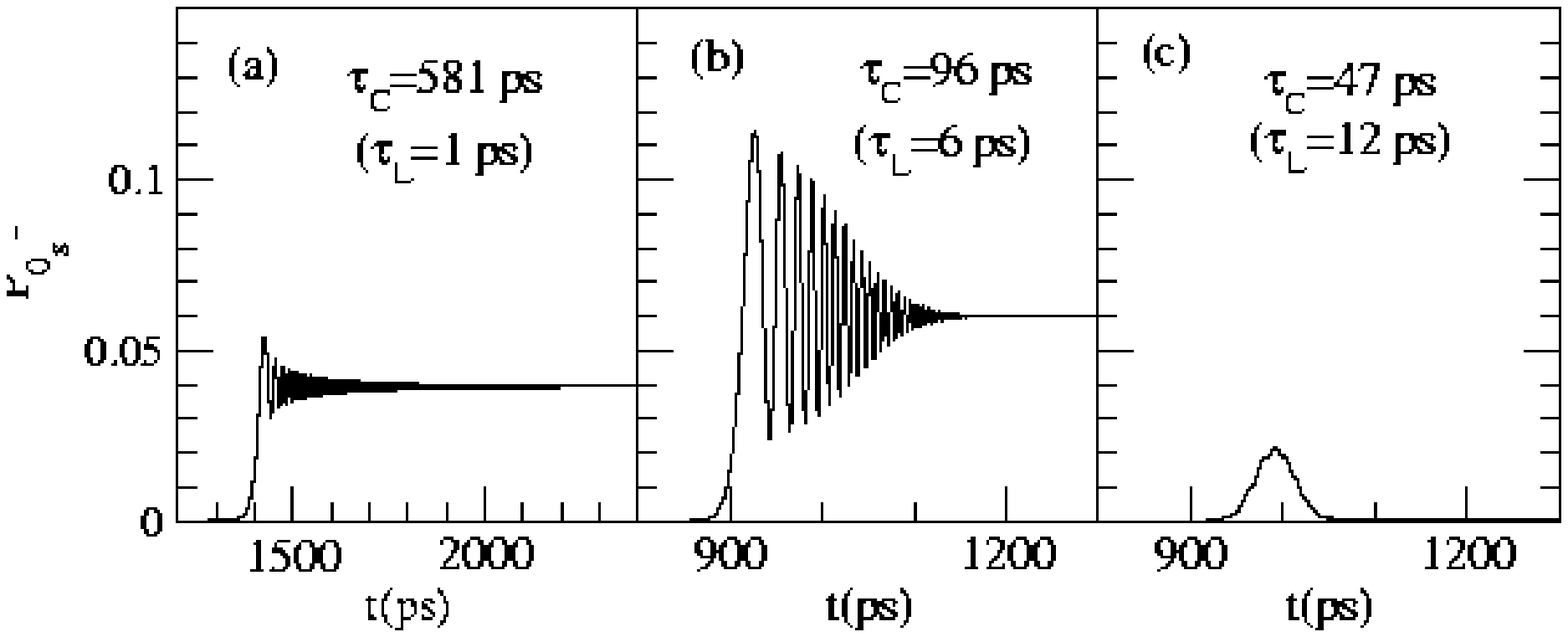}
}
\caption{Time evolution of the $0_g^-$ population (for a total population normalized at 1 on the grid)  during the photoassociation process, by using the ``really chirped'' pulses  whose envelopes f(t) are represented in Fig. \ref{fig:pulses-env}b,c): a) $0_g^-(6s,6p_{3/2})$ population for a pulse with $\tau_C$= 581 ps ($\tau_L$=1 ps), centered at $t_P=1500$ ps, b) $0_g^-(6s,6p_{3/2})$ population for a pulse with $\tau_C$= 96 ps ($\tau_L$=6 ps), centered at $t_P=1000$ ps, c) $0_g^-(6s,6p_{3/2})$ population for a pulse with $\tau_C$= 47 ps ($\tau_L$=12 ps), centered at $t_P=1000$ ps.}
\label{fig:pop0g-c3}
\end{figure}
\begin{figure*}
\resizebox{0.75\textwidth}{!}{%
\includegraphics{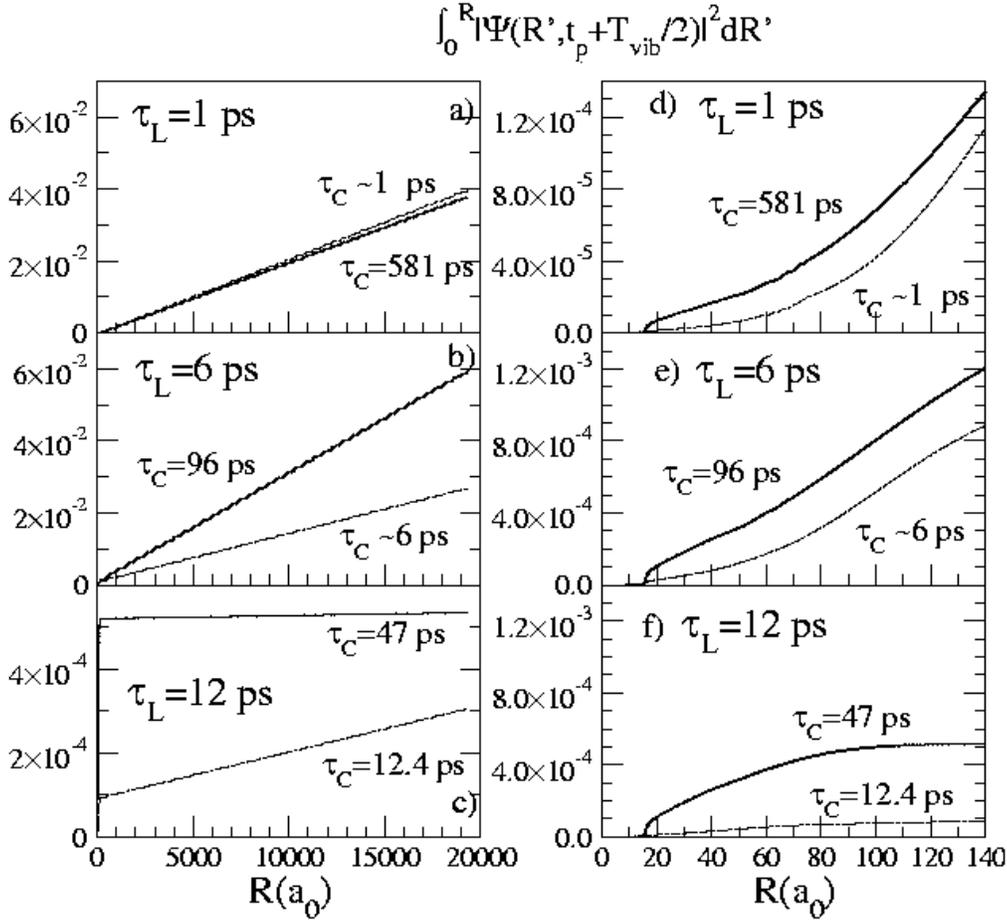}
}
\caption{Distribution of the $0_g^-$ radialy integrated population $P_{0_g^-}(R,t_{foc})$ at
 $t_{foc}=t_P+\frac{T_{vib}}{2}=t_P$+130 ps, for ``really-chirped'' and ``almost no-chirped'' pulses. Results obtained with pulses having the same $\tau_L$ (the same spectral width $\hbar \delta \omega$) are represented together: thick lines for ``really-chirped'' pulses ($\tau_C \gg \tau_L$), and thin lines for ``almost no-chirped'' pulses ($\tau_C \approx \tau_L$). Left column: distribution of the $0_g^-$ population until $L_R$=19250 a$_0$. Right column: population until  R=140 a$_0$. ``Really-chirped'' pulses give a much larger population transfer in the $0_g^-$ state, as the spanned energy range  2$\hbar |\chi| \tau_C$ is much larger; they give also much more population at small distances. (The results show the  population in the $0_g^-$ channel relative to a total population normalized at 1 on the grid.)}
\label{fig:pR-0.5Tvib-3c+-}
\end{figure*}

The Figures \ref{fig:pop0g-nc3} and  \ref{fig:pop0g-c3} show the time evolution of the population $P_{0_g^-}(t)$ transferred in the $0_g^-(6s,6p_{3/2})$ state during the photoassociation process, using the ``almost no-chirped''  pulses with $\tau_C \approx \tau_L$=1, 6, 12 ps, and  the ``really chirped'' pulses with $\tau_C$= 581, 96, 47 ps, respectively. On the other hand, in Fig. \ref{fig:pR-0.5Tvib-3c+-} is represented $P_{0_g^-}(R,t_{foc})$,  the  $0_g^-(6s,6p_{3/2})$ radialy integrated population as a fonction of the distance R, at the moment $t_{foc}=t_P+\frac{T_{vib}}{2}=t_P$+130 ps, for all the pulses. Results for pulses having the same $\tau_L$ are represented together, the left column showing results until $L_R=19250$ a$_0$ (the limit of the spatial grid), and the right column showing the repartitions of the same  populations at small distances, until $R=140$ a$_0$.

At a first view the time evolutions of the $0_g^-$ population given by the three classes of pulses, with $\tau_L$=1, 6, 12 ps (and with three increasing energies of the pulse $E_{pulse}$)  are quite different, and this can be understood looking at the  characteristic spectral bandwidths (see the Table \ref{tab:pulses}) and comparing them with the detuning.

 As it was mentioned, the pulses with $\tau_L$=1 ps  will massively populate the  $0_g^-$ continuum (see Figs.  \ref{fig:pop0g-nc3}a) and \ref{fig:pop0g-c3}a)), leading at the end to the same large value of the $0_g^-$ population ($P_{0_g^-}$=0.04). In this case,  the spectral width $\hbar \delta \omega$ and the pulse energy $E_{max}$ appear as the only parameters playing in the results, independently on the energy range 2$\hbar|\chi| \tau_C$ swept during the pulse. The distributions of the $0_g^-$ population function of the distance R are nearly identical for both pulses (see Fig. \ref{fig:pR-0.5Tvib-3c+-}a) ), except at short distances (Fig. \ref{fig:pR-0.5Tvib-3c+-}d)). It appears that due to the high value of the spectral width $\hbar\delta \omega > \delta_L^{at}$, the population is mainly transferred at  very large distances R $> 1000$ a$_0$ (non resonantly for the ``almost no-chirped'' pulse with  $\tau_C \approx \tau_L$=1 ps).  Only a little amount is excited at smaller distances
 $R<100$ a$_0$, where the chirped pulse with $\tau_C=581$ ps has a large ``resonance window'' (large value $\hbar |\chi| \tau_C$) and it is more efficient than the short pulse of 
$\tau_C \approx$ 1 ps, with a narrow ``resonance window''. It is interesting to remark that, for the pulse with  $\tau_C \approx$ 1 ps the population transfer is adiabatic ($\alpha_{max}>1$) during the time window $\lbrack -\tau_C,\tau_C \rbrack$, but the adiabaticity condition cannot be satisfied for the very long pulse with $\tau_C=581$ ps, for which $\alpha_{max}$ is not defined (see Table 1). In fact, the oscillations which can be observed in the time evolution of the $0_g^-$ population in Fig. \ref{fig:pop0g-c3}a) correspond to a strong Rabi coupling at very large distances (described by the Eq. (\ref{eq:Rabicoupl})) and are the signature of a  strong nonadiabatic behaviour in the population transfer. This non adiabatic population transfer at the large distances swept by the instantaneous crossing point results in a large population remaining in the $0_g^-$ surface after the end of the pulse. Similar oscillations indicating a nonadiabatic transfer can be equally observed in Fig. \ref{fig:pop0g-c3}b), in the evolution of the $0_g^-$ population during the excitation with the pulse with $\tau_C=96$ ps ($\tau_L$= 6 ps), to whom corresponds a small value of $\alpha_{max}=0.5$.

The two other classes of pulses, with $\tau_L$= 6 and 12 ps, have much narrow spectral widths, smaller than the detuning $\delta_L^{at}$=2.656 cm$^{-1}$. For these cases, it appears clearly that, for the same $\tau_L$, the ``really chirped'' pulse  is much more efficient for the $0_g^-$ population transfer, both for the total transfer, as for the transfer at small distances (see Figs. \ref{fig:pR-0.5Tvib-3c+-}b,c,e,f). The radially integrated population $P_{0_g^-}(R,t_{foc})$ is generally a linearly increasing function of the distance $R$ 
(see Fig. \ref{fig:pR-0.5Tvib-3c+-} left column) except for the pulse with $\tau_C$= 47 ps ($\tau_L$= 12 ps), for which $P_{0_g^-}(R,t_{foc})$ reaches its limiting value at small $R \approx 140$ a$_0$
(Fig. \ref{fig:pR-0.5Tvib-3c+-}c,f)), because the population transfer takes place adiabatically in a ``photoassociation window'' \cite{eluc04}. Indeed, in this case, the energy range  resonantly swept by the pulse to the large distances during the period $\lbrack -\tau_C,\tau_C \rbrack$ is $\hbar |\chi| \tau_C=1.17$ cm$^{-1}$, the spectral width is $\hbar \delta \omega=$1.23 cm$^{-1}$ (satisfying $\hbar |\chi| \tau_C+\hbar \delta \omega < \delta_L^{at}$)  and the population transfer keeps an adiabatic character ($\alpha_{max}=0.87$) both in the resonance window, and at large internuclear distances ($\Delta(R) \sim\delta_L^{at}$ for $R>200$  a$_0$).

 On the contrary, the chirped pulse with $\tau_C$= 96 ps ($\tau_L$= 6 ps) does not produce excitation in a photoassociation window, but everywhere at large distances, because the energy range swept during the time window (to the large distances) approaches the dissociation limit ($\hbar |\chi| \tau_C=2.41$ cm$^{-1}$ $\approx$ $\delta_L^{at}$). 
\begin{figure}
\resizebox{0.45\textwidth}{!}{%
\includegraphics{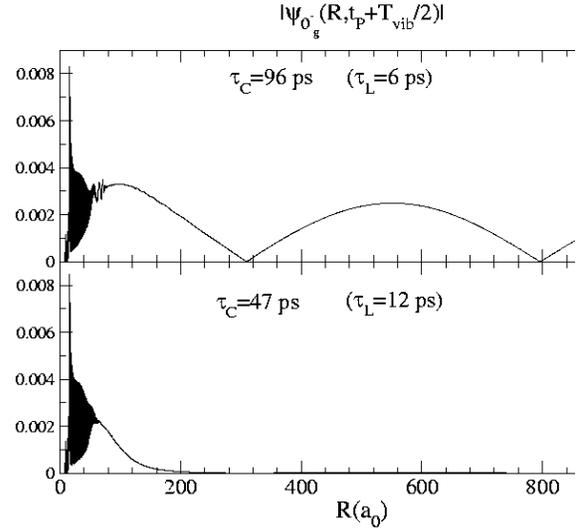}
}
\caption{ Excited wavepackets localized at $t_{foc}=t_P+\frac{T_{vib}(v_0)}{2}$ at the inner turning point of the outer well of the $0_g^-$ potential. Comparison of two typical final states: a) transfer at large distances, and b) excitation in a ``photoassociation window''.}
\label{fig:wf-0.5tvib}
\end{figure}

We have to remark that the very different evolutions of the $0_g^-$ populations during the pulses with $\tau_L$= 6 ps (see Figs. \ref{fig:pop0g-nc3}b), \ref{fig:pop0g-c3}b)), on the one hand, and with $\tau_L$= 12 ps (see Figs. \ref{fig:pop0g-nc3}c), \ref{fig:pop0g-c3}c)) on the other hand, are typically for two kinds of results that can be obtained: at the end of the pulse one can have either population excited everywhere at large distances (in an amount increasing with R, reflecting the amplitude of the initial continuum), or, if the spectral band of the pulse is narrow enough, one can observe a selective adiabatic transfer of population in a certain domain of distances, representing a ``photoassociation window'', and very small population transfer outside the window (where the transfer is also adiabatic, but where there are no instantaneous crossings, which explains the small value of the remaining $P_{0_g^-}$ at  the end of the pulse). An illustration of these results is made in Fig. \ref{fig:wf-0.5tvib} showing the $0_g^-$ wavepackets at $t_{foc}=t_P+\frac{T_{vib}}{2}=t_P+$ 130 ps, a) for the chirped pulse with $\tau_C$= 96 ps ($\tau_L$= 6 ps), giving population transfer on the whole spatial grid, and b) for the chirped pulse with $\tau_C$= 47 ps ($\tau_L$= 12 ps), producing transfer in a ``photoassociation window''. Due to the choice of the $\chi$ value, at $t_{foc}=t_P+\frac{T_{vib}(v_0)}{2}$ both wavepackets are focussed at the inner turning point of the vibrational eigenstate $v_0$ excited at $t=t_P$.

We are especially interested in pulses producing the maximum transfer of $0_g^-$ population to the small and intermediate distances (see Fig.\ref{fig:pR-0.5Tvib-3c+-} for this analysis).  From this point of view, the pulses with $\tau_L$= 6 ps ($\tau_C$= 96 ps and $\tau_C$= 6.012 ps) and the pulse with $\tau_C$= 47 ps ($\tau_L$= 12 ps) are giving the bigger population at small R. Their results can be seen and compared in Figs. \ref{fig:pR-0.5Tvib-3c+-}e),f). At $t_{foc}=t_P+\frac{T_{vib}}{2}$, both pulses with  $\tau_C$= 6.012 ps  ($\tau_L$= 6 ps) and with $\tau_C$= 47 ps ($\tau_L$= 12 ps) give $P_{0_g^-}$(100 a$_0$, $t_{foc}$) $\approx 4 \times 10^{-4}$, but the pulse with $\tau_C$= 47 ps, which produces a ``photoassociation window'', is more favorable to transfer population at smaller R-values $R<100$ a$_0$. Nevertheless, it is the pulse with $\tau_C$= 96 ps ($\tau_L$= 6 ps) which seems to be the most efficient for the $0_g^-$ population transfer, at large distances, but also to the inner region ($P_{0_g^-}$(100 a$_0$, $t_{foc}$) $\approx 8 \times 10^{-4}$),  because even after the end of the pulse, the population will be accelerated inside from the large distances.

Let's remark that, as dicussed in Ref.\cite{cao00},  the  negative chirp has a noticeable contribution in the acceleration of the wavepacket created in the excited state $0_g^-$ towards short internuclear distances. Indeed, $\omega(t)$ decreasing with time, the instantaneous crossing point moves to smaller distances, following the motion of the wavepacket which is accelerated inside the potential well. This is favorable to our goal of maximizing $0_g^-$ population at small distances. On the other hand, this means that an interpretation of the transfer process within the impulsive approximation  could easily be invalidated. For example, in our case, the population in the initial continuum state integrated until $R=100$ a$_0$ is:
\begin{equation}
\int_{0}^{R=100} | \Psi_{\Sigma,T=54 \mu K}(R') |^{2} dR'= 3 \times 10^{-4}
\label{eq:psigma}
\end{equation}
and this is what would be the maximum value of the integral $\int_{0}^{R=100} | \Psi_{0_g^-}(R',t_{foc}) |^{2} dR'$ if the impulsive approximation would be valid until $t_{foc}$. But we have shown that, in the cases of some of the pulses discussed before, (with $\tau_C \approx$6 ps, $\tau_C$= 96 ps and $\tau_C$= 47 ps),  $P_{0_g^-}$(100 a$_0$, $t_{foc}$) reaches indeed bigger values, proving the acceleration towards shorter internuclear distances for times $t<t_{foc}$.

\subsection{Population in the last vibrational levels of the ground state} 
In the non perturbative regime there is a redistribution of population in the ground surface $a^3\Sigma_u^+(6s,6s)$, arising as well in the bound spectrum, as in the dissociation continuum. There appears a ``hole'' in the initial wavefunction around $R_L$,  a large part of this population being transferred in  bound vibrational levels of $0_g^-$ and $a^3\Sigma_u^+$.

\begin{figure}
\resizebox{0.5\textwidth}{!}{%
\includegraphics{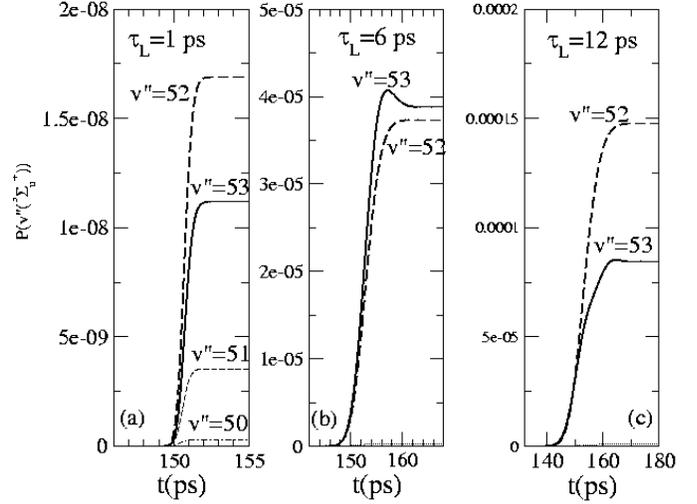}
}
\caption{Populations in the last vibrational levels v''($a^3\Sigma_u^+(6s,6s)$) of the ground state, for ``almost no-chirped'' pulses ($\tau_C \approx \tau_L$). (The total population is normalized at 1 on the grid.)}
\label{fig:popsigma-nc3}
\end{figure}

\begin{figure}
\resizebox{0.5\textwidth}{!}{%
\includegraphics{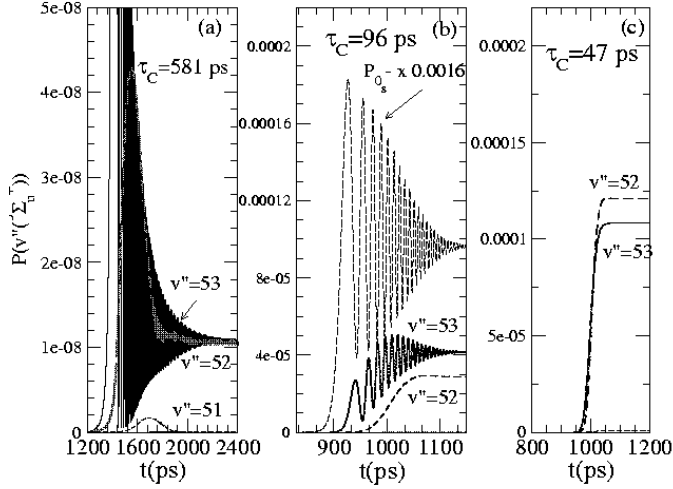}
}
\caption{Populations in the last vibrational levels v''($a^3\Sigma_u^+(6s,6s)$) of the ground state for chirped pulses with $\tau_C$= 581 ps, 96 ps, and 47 ps. (The total population is normalized at 1 on the grid.) }
\label{fig:popsigma-c3}
\end{figure}

Due to the coupling at large distances between the two electronic states, the last vibrational levels of the $a^3\Sigma_u^+(6s,6s)$ state are noticeable populated during the photoassociation process.
 Figs. \ref{fig:popsigma-nc3} and \ref{fig:popsigma-c3} show the evolution of the population in these last levels ($v''=53$ is the last vibrational level \cite{eluc04}). If one compares pulses with the same $\tau_L$,  the final transferred population is roughly of the same order, but for $\tau_L$= 1 or 12 ps it is bigger for ``almost no-chirped'' pulses, probably because the maximum coupling W$_{max}$ is more intense for shorter pulses. The pulses with $\tau_L$= 6 ps ($\tau_C$= 6.012 ps and 96 ps) give almost the same population of v''=52, 53.
In Fig. \ref{fig:popsigma-c3}b) we show the Rabi oscillations in phase opposition between the $0_g^-$ population $P_{0_g^-}$ (which is distributed on the whole spatial grid of $L_R$=19250 a$_0$) and the population of the last vibrational state of the $a^3\Sigma_u^+(6s,6s)$, with v''= 53, whose wavefunction extends at very large distances (the last oscillation is between 150 and 1200 a$_0$, see Fig. \ref{fig:art2-init}a)). This result is symptomatic for the  efficient exchange of population between the two channels at very large distances, already emphasized, and due to the small detuning and large overlap integral (see Appendix C, Fig. \ref{fig:recouvr}). Around $t_p=1000$ ps, the period of the observed Rabi oscillations is about 12.4 ps, in agreement with the result given by the formula (\ref{eq:Rabicoupl}). The oscillations which can be observed  in Figs. \ref{fig:popsigma-c3}a) and b) are characteristic for nonadiabatic effects in the transfer of populations.

 One has to emphasize that, in the cases of adiabatic transfer in a limited spatial window  (exemplified here with the pulse  $\tau_C$= 47 ps, and in Ref. \cite{eluc04} with a pulse having $\tau_C$= 34.8 ps), the populations in the last bound states v''= 52,53 of the ground potential at the end of the pulse is of the same order as the population $P_{0_g^-}$  remained in the $0_g^-$ levels resonantly excited by the pulse in the so-called ``photoassociation window''. For example, for the pulse with  $\tau_C$= 47 ps, $P_{0_g^-}(E_0)=5.34 \times 10^{-4}$ (see Table 1), and the population $P_{^3\Sigma_u}(v''=52) +P_{^3\Sigma_u}(v''=53) \approx 2.25 \times 10^{-4}$ (see Fig. \ref{fig:popsigma-c3}c) ). Then, in this case, immediately after the pulse, the individual $0_g^-$ bound levels belonging to the ``photoassociation window'' are less populated than these last  $a^3\Sigma_u^+(6s,6s)$ states. But it has to be noted that v''=52,53 have wavefunctions extending at very large distances (hundreds of a$_0$), being then well populated during the exchange of population between the two channels, favoured by the good overlap at large distances, as explained before. In exchange, the levels $0_g^-$  populated in a ``photoassociation window'' are bound at much smaller distances, as their outer turning points are between $R_{min}$=84 a$_0$ and $R_{max}$=117 a$_0$.

\subsection{Evolution of the $0_g^-$ wavepacket after the pulse}
We shall briefly describe the wavepackets evolution  in the $0_g^-$ surface after the pulse, in the two typical cases discussed before: excitation in a limited spatial ``photoassociation  window'' and excitation at all distances, with massive transfer of population at large distances. Obviously, these two types of wavepackets created by the photoassociating  pulse are extremely different, and their dynamics is significant in view of some anticipation of the  results that could be brought by a second pulse. This second pulse can be even identical with the first one, but time-delayed (in the analysis of the repetition rate of the laser) or a different pulse, if the goal is the stabilization of the system by stimulated emission to low vibrational levels of the ground state. These subjects will be treated in a future article.

\subsubsection{Excitation in ``a window'': Vibrational dynamics. }
\begin{figure}
\resizebox{0.45\textwidth}{!}{%
\includegraphics{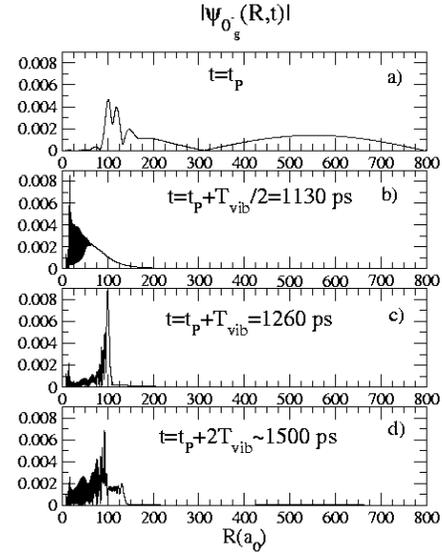}
}
\caption{$0_g^-$ wavepackets evolution after the pulse with  $\tau_C=47$ ps ($\tau_L=12$ ps).}
\label{fig:wf12-t}
\end{figure}
Fig. \ref{fig:wf12-t} shows the evolution of the  $0_g^-$ wavepacket adiabatically excited ($\alpha_{max}=0.87)$ in a limited spatial window, with a chirped pulse of time width  $\tau_C=47$ ps ($\tau_L=12$ ps). In fact, during the pulse, a large amount of population is transferred to levels close to the Cs$_2$(6s+6p$_{3/2}$)  dissociation limit (see the wavepacket at $t=t_P$, in Fig. \ref{fig:wf12-t}a) ), but, due to the adiabaticity of the population transfer outside the ``photoassociation window'', this population goes back to the ground state and only a small range of distances (R$<$200 a$_0$) remains populated. Then, after the pulse, one can observe a typical vibrational dynamics, as it can be seen in the figure for some selected moments: at $t=t_P+\frac{T_{vib}(v_0)}{2}$ the packet is focalized at the inner turning point of $v_0$, at $t=t_P+T_{vib}(v_0)$ the packet comes back at the external turning point of $v_0$ ($R_L=93.7 \ a_0$),  at $t=t_P+2T_{vib}(v_0) \approx t_P+ 500$ ps, one can distinguish two parts: one, with a maximum around $R_L=93.7 \ a_0$, is composed by  levels with vibrational periods close to $T_{vib}(v_0)$; the other has a maximum around $R=130 \ a_0$: indeed, the vibrational levels of the $0_g^-$  potential having the  external turning point around this distance vibrate with $T_{vib} \approx 500$ ps.

\subsubsection{Excitation at large distances: acceleration to the inner region}
\begin{figure*}
\resizebox{0.75\textwidth}{!}{%
\includegraphics{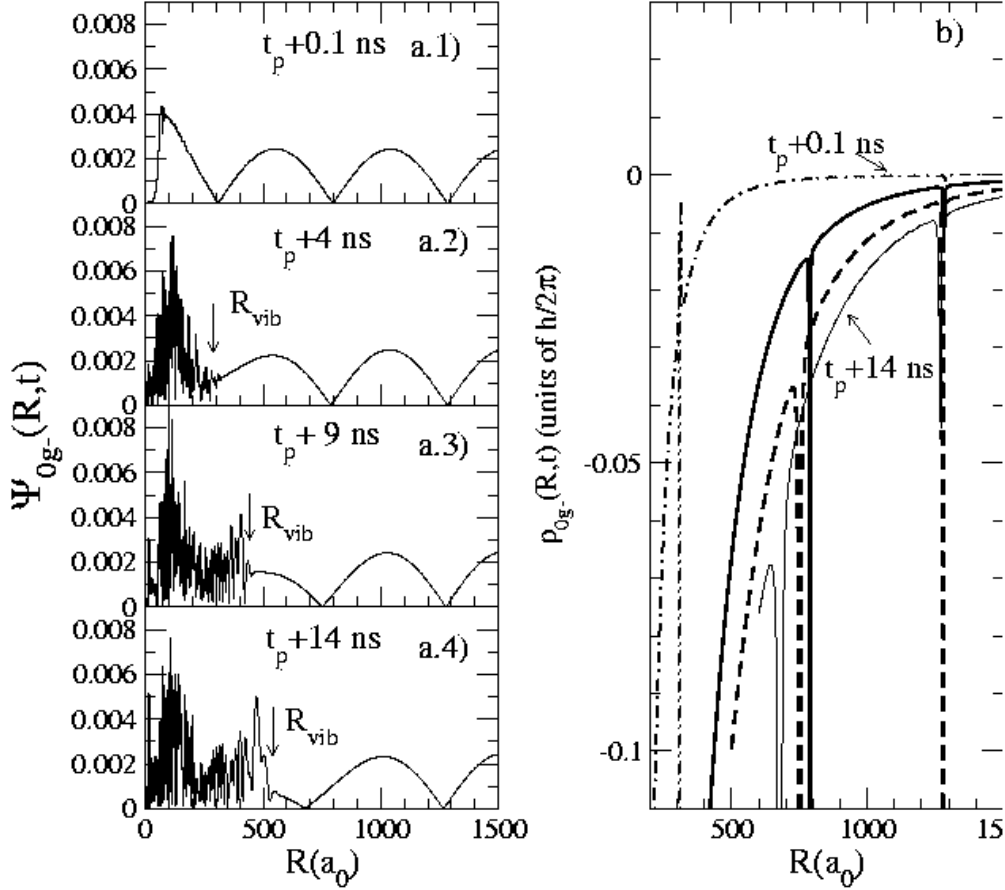}
}
\caption{a.1-4) $0_g^-$ wavepackets evolution after the pulse with  $\tau_C=96$ ps ($\tau_L=6$ ps), at different moments after the end of the pulse (whose maximum is at $t_P=1000$ ps): t= $t_P$+0.1 ns, $t_P$+ 4 ns, $t_P$+9 ns, $t_P$+14 ns. b) The R-variation of the  momentum $\frac{1}{\hbar}p_{0_g^-}(R,t)$ at the same times: $t_P$+0.1 ns (dot-dashed line), $t_P$+ 4 ns (thick continuum line), $t_P$+9 ns (dashed line), $t_P$+14 ns (thin continuum line), showing the acceleration of the wavepacket to the inner region. The singularities in the momentum behaviour are due to the definition modulo $\pi$ of the phase. }
\label{fig:wflux}
\end{figure*}

For pulses having  sufficiently large bandwidths ($\delta \omega > \delta_L^{at}$), the population transferred at large distances during the pulse remains on the excited surface $0_g^-$ also after the end of the pulse. Just after the end of the pulse, the radial distribution $| \Psi_{0_g^-}(R,t) |^{2}$ of the probability density reproduces that of the initial collisional state on the  $^3\Sigma_u^+$ surface at sufficiently large distances $R>500$ a$_0$. After that, the wavepacket evolves in the $-C_3/R^3$ potential and it is accelerated toward the inner region. Fig. \ref{fig:wflux} describes the evolution of the wavepacket created by the pulse with  $\tau_C=96$ ps ($\tau_L=6$ ps), at different moments $t_P+\Delta t$, with $\Delta t=$0.1, 4, 9, and 14 ns. As it can be seen in Fig. \ref{fig:wflux}a), for $R \ge 1500$  a$_0$, there is no noticeable modification of the $| \Psi_{0_g^-}(R,t) |$, even for the largest value of $\Delta t$. As $\Delta t$ increases, the nodes occuring at $R \ge 500$  a$_0$ ($R_1 \approx 800$ a$_0$ and  $R_2 \approx 1270$ a$_0$, for example) begin to be shifted toward small R values. In the range $R \le 500$  a$_0$, vibrational motion can be  observed winning progressively distances $R_{vib}(t_v)$ in the outer well of the $0_g^-$  potential, after a time $t_v$ approaching  the vibrational period $T_{vib} \approx t_v$ of the levels  whose wavefunctions have the outer turning points around $R_{vib}(T_{vib})$. For example, for v=162, with $T_{vib}\approx 13.7 $ ns, $R_{vib}=490$ a$_0$. (The largest time $\tau_{max}=2.9$ $\mu s$ relevant for the presently used grid with  $L_R=19250$ a$_0$ correspond to vibrational movement at distances $R \approx 4000$ a$_0$.) 

The acceleration of the $0_g^-$ wavepacket can be estimated introducing  the radial flux \cite{messiah}:
\begin{equation}
J_{0_g^-}(R,t)=-\frac{i\hbar}{ \mu } Im [ \Psi^{\ast}_{0_g^-}(R,t) \frac{\partial}{\partial R} \Psi_{0_g^-}(R,t)],
\end{equation}
with $\mu=121136$ a. u. the reduced mass of Cs$_2$. Writing the wavepacket as $\Psi_{0_g^-}(R,t)=| \Psi_{0_g^-}(R,t) | e^{i \phi_{0_g^-}(R,t)} $, one has $\frac{1}{\hbar}p_{0_g^-}(R,t)=\frac{\mu}{\hbar}\frac{J_{0_g^-}(R,t)}{| \Psi_{0_g^-}(R,t) |^{2}}=\frac{\partial}{\partial R} \phi_{0_g^-}(R,t)$. Fig. \ref{fig:wflux}b) reports the R-variation of the  momentum $\frac{1}{\hbar}p_{0_g^-}$ at the times $t_p+\Delta t$;  the rapid increase at intermediate R values $200 \ a_0 \le R \le 700 \ a_0 $, for $0.1 \ ns < \Delta t < 5 \ ns$ is clearly illustrated. This increase in the momentum transfer during the pulse is a signature of the ``kick'' given by the laser light to the molecule.  For ``really chirped'' pulses the photoassociation probability is relatively important, and the population  transferred at intermediate distance is accelerated to the inner region a short time after the end of the pulse, increasing significantly the population at short  distances. Optimization of the pulse parameters in order to give to the population  transferred at intermediate distances a strong momentum to the inner region will be considered in the future. 

\section{The Photoassociation probability from a thermal  average over  the incident kinetic energies}
Supposing thermal equilibrium at the temperature T, the initial density matrix can be expressed in terms of energy-normalized collisional eigenstates of energies E in the ground state $a^3\Sigma_u^+$, as in the formula (\ref{eq:averen}). Then the total probability per pump pulse that a given pair of atoms to be photoassociated into the excited state $0_g^-$ at the temperature T, is:
\begin{equation}
{\cal P}_{0_g^-}(T)=\frac{1}{Z} \int_0^{\infty} dE e^{-\beta E}\bar{P}_{0_g^-}(E),
\label{eq:paaverage}
\end{equation}
where $\bar{P}_{0_g^-}(E)$ accounts for the density of probability (probability per unit energy range) that an energy normalized continuum state of energy E belonging to the ground state $a^3\Sigma_u^+ $ to be photoassociated at the end of the pulse in the excited state  $0_g^-$. $\bar{P}_{0_g^-}(E)$ has dimensions of $1/E$, being obtained with the formula (\ref{eq:normen}) from the dimensionless photoassociation probability $P_{0_g^-}(E)$ in the  $0_g^-$ state:
\begin{equation}
\bar{P}_{0_g^-}(E)=\{ \frac{\partial E}{\partial n}\}^{-1} P_{0_g^-}(E),
\end{equation}
We shall discuss in the following the evaluation of the integral (\ref{eq:paaverage}).

\subsection{Analytical thermal average using the scaling law in the vicinity of $E \to 0$}
In Ref. \cite{eluc04} and in the most part of this paper we present results of photoassociation calculations considering only the $s$-wave and taking as the initial state in the $a^3\Sigma_u^+(6s,6s)$ ground potential a single continuum state of the thermal energy distribution, with the energy $E_0$.  In our calculation $E_0=1.7211 \times 10^{-10}$ a.u.$=3.778 \times 10^{-5}$ cm$^{-1}$,  being, in our method, the energy of a continuum state belonging to the discretized continuum which is calculated numerically with the Sine Mapped Grid method in a box of radius $L_R= 19250$ a$_0$.  Then, with this choice of the initial continuum state, the temperature in the thermal ensemble of atoms is arbitrarily defined by  $k_B T=E_0$, as being  $T = 54$ $\mu K$. At this temperature, the initial Boltzmann distribution (see the formula (\ref{eq:averen}) ) is narrow ($e^{-3}=0.05$), and a first approximation is to represent it by its mean energy $E_0$, which is justified by the fact that the spectral widths $\hbar \delta \omega$ of the  pulses considered in photoassociation (see Table 1) are much larger than $k_B T$. Then, the integral (\ref{eq:paaverage}) can be evaluated  from  $\bar{P}_{0_g^-}(E_0)$, by using a threshold scaling law 
to estimate the probabilities $\bar{P}_{0_g^-}(E)$ for $E \ne E_0$   \cite{mackholm94}. 

We shall discuss the conditions making valid the use of the scaling law for estimations of the photoassociation rates. We  emphasize that such a discussion makes sense only for the photoassociation with a cw-laser or with a pulse leading to  the formation of a spatially localized wavepacket on the excited surface \cite{mackholm94,eluc04}, giving what we have called a ``photoassociation window''.  In such a case, the main contribution to the photoassociation process towards high excited vibrational levels is provided by the very localized range of the internuclear distances $[R_{min},R_{max}]$, around $R \sim R_L$, swept by the instantaneous crossing point $R_c(t)$ during the time window $[-\tau_{c},\tau_{c}]$ \cite{eluc04}.  The photoassociation yield is determined  by the overlap between the initial stationary continuum $|^3\Sigma_u^+, E>$ in the ground state and the excited wavefunctions $|0_g^- , v>$ (see Appendix C, Fig. \ref{fig:recouvr}).

A scaling law for the behaviour of the continuum wavefunctions can be obtained from their asymptotic forms.  For a very  small detuning $\delta_L^{at}$, the crossing point $R_L$ of the dressed potentials is at a  distance large enough making that,  even at a low continuum energy E,  the potential in $R_L$ can be considered as negligibile: $E > \frac{C_6}{R_L^6}$. Then, for a $s$-wave,  the continuum wavefunction can be described by its asymptotic behaviour \cite{julienne96}:
\begin{equation}
|\Psi_{g,E}(R)| \approx \sqrt{\frac{2 \mu}{\pi \hbar^2}}\frac{\sin[k(R-L)]}{\sqrt{k}}
\label{eq:asymptwf} 
\end{equation}
$L$ is the scattering length of the ground surface. This means that, for  sufficiently low collision energies:
\begin{equation}
 E \ll \frac{1}{2 \mu}|\frac{\pi \hbar}{R_L-L }|^2=k_B T_{an}
\label{eq:smalle}
\end{equation}
one obtains the following probability density in $R_L$:
\begin{equation}
|\Psi_{g,E}(R_L)|^2 \approx\frac{2 \mu}{\pi \hbar^2} k(R_L-L)^2 \sim \sqrt{E}
\end{equation}
as $E=(\hbar k)^2/2 \mu$.
Then, for pulses producing a spatially localized wavepacket on the excited surface, and for sufficiently small energies, one can assume the following law for the probabilities $\bar{P}_{0_g^-}(E) \sim |\Psi_{g,E}(R_L)|^2 $:
\begin{equation}
\bar{P}_{0_g^-}(E)=\bar{P}_{0_g^-}(E_0) \sqrt{ E/E_0}
\label{eq:scal}
\end{equation}
Using the relation (\ref{eq:scal}), the integral (\ref{eq:paaverage})  can be  evaluated as being:
\begin{equation}
{\cal P}^{an}_{0_g^-}(T)=\frac{1}{Z} \bar{P}_{0_g^-}(E_0) \frac{\sqrt{\pi}}{2\sqrt{E_0} }( k_B T)^{3/2}
\label{eq:pae0}
\end{equation}
$Z{\cal P}^{an}_{0_g^-}(T)$ represents the probability that a pair of atoms, belonging to a gas in a volume V and at the temperature T, and described only by $s$-waves,  to be photoassociated in the  $0_g^-(6s+6p_{3/2})$ state. 

For a typical number which can be obtained for this quantity,  we shall take as example  the pulse studied in great detail in Ref.\cite{eluc04}, whose characteristics  are given in the Table 1, and having the duration $\tau_C=34.8$ ps ($\tau_L=15$ ps).   The induced dynamics (which is very similar to that resulting from the ``really chirped'' pulse  with $\tau_C=46.6$ ps and $\tau_L=12$ ps) results in an adiabatic population transfer in a spatial  range with $R_{min}=85$ a$_0$, $R_{max}=110$ a$_0$, with $\alpha_{max}=0.95$, and populating about 15 vibrational levels in the vicinity of $v_0=98$ level in the external well of the $0_g^-(6s+6p_{3/2})$ potential. It corresponds to the photoassociation probability $P_{0_g^-}(E_0)=3.245 \times 10^{-4}$. Then, taking into account the density of states $\frac{dn}{dE}|_{E_0}=1.1415 \times 10^{11}$ a.u. at $E_0/k_B=54.35$ $\mu$K, for a gas of cesium atoms at  $T=54$ $\mu$K we have used the formula (\ref{eq:pae0}) to obtain:
\begin{equation} 
Z{\cal P}^{an}_{0_g^-}(T=54 \ \mu K, \tau_C=34.8 \  ps )=0.00560
\label{eq:pa1}
\end{equation}
 with $\tau_C$ characterizing the pulse used for photoassociation.

\subsection{Average implicitely accounting for real threshold effects}

As discussed in a previous section, in the conditions of temperature and detuning discussed in the present paper, the asymptotic behaviour of the continuum wavefunction having the energy $E_0=k_B T$, $T = 54$ $\mu K$, is not reached at $R=R_L=94$ a$_0$. Indeed, the potential energy strongly determines the structure of the initial wavefunction for  $R \leq R_N=82.3$ a$_0$, $R_N$ being the position of the last common node. This observation is in agreement with the discussion of Ref.\cite{julienne96}, showing that the asymtotic behaviour in a $-\frac{C_6}{R^6}$ potential is reached for $R \gg R_B= (\frac {\mu C_6}{10 \hbar^2 })^{1/4}$, giving $R_B=95$ a$_0$ for the $^3\Sigma_u^+(6s,6s)$ potential.

Then, in the present studied example of cold atoms photoassociation, the detuning  $\delta_L^{at}=2.656$ cm$^{-1}$ is too large ($R_L \approx R_N$), and the temperature $T = 54$ $\mu K$ not sufficiently small compared to $T_{an} = 69.2$ $\mu K$ (see the relation (\ref{eq:smalle}) ), to allow the evaluation of the photoassociation rate from an analytical thermal average. The scaling law in  $\sqrt{ E}$ being not valid, it is necessary to explicitely study the energy variation of  $\bar{P}_{0_g^-}(E)$, by considering different initial collisional states of energy E in the $^3\Sigma_u^+$ potential, and having a node at $L_R$. In this case, the real threshold effects are completely and correctly accounted for implicitely by performing numerical integration in Eq. (\ref{eq:paaverage}). For example, for the pulse with $\tau_C=34.8$ ps ($\tau_L=15$ ps) studied in  Ref.\cite{eluc04}, we have performed 21 time-propagation calculations for collisional energies in the range $36.6 \ nK < E/k_B < 633 \ \mu K$. Let us emphasize that this detailed analysis of the threshold effects has become possible owing to the Mapped Sine Grid method \cite{willner04} for which a large box of dimension $L_R$ can be considered. The energy-dependence  of $\bar{P}_{0_g^-}(E)$ will be analyzed in a further publication \cite{elianeanne}. It differs strongly from the  $\sqrt{ E}$ scalling law. $\bar{P}_{0_g^-}(E)$ increases very rapidly at threshold exhibiting a very sharp asymetrical resonance like structure  with a maximum at 
$ E/k_B \approx 7.9 \ \mu K$ (smaller than the temperature studied presently) with a FWHM $\Delta E/k_B \approx 4.7 \ \mu K$.
For the temperature  $T=54$ $\mu$K, numerical integration in Eq. (\ref{eq:paaverage}) leads to 
\begin{equation} 
Z{\cal P}_{0_g^-}(T=54 \ \mu K, \tau_C=34.8 \  ps )=0.00685
\label{eq:pa2}
\end{equation}
The rather good agreement between the analytic average  (\ref{eq:pa1}) calculated with the scaling law, and the average (\ref{eq:pa2}) containing the threshold effects is completely fortuitous.

The two pulses described by the same parameters $\delta_L^{at}$, $W_L$ and $\chi$, and differing only by their durations 
$\tau_C=34.8$ ps ($\tau_L=15$ ps) and $\tau_C=46.6$ ps ($\tau_L=12$ ps) are associated with rather similar ``photoassociation windows'' characterized by the values ($\alpha_{max}=0.95$, $\hbar |\chi| \tau_C=0.87$ cm$^{-1}$) and ($\alpha_{max}=0.87$, $\hbar |\chi| \tau_C=1.17$ cm$^{-1}$). For the same initial collisional state $E_0/k_B=54.35$ $\mu$K, the corresponding photoassociation probabilities are  $P_{0_g^-}(E_0)=3.245 \times 10^{-4}$ and $5.340 \times 10^{-4}$. The photoassociation probability is larger for the longer pulse with $\tau_C=46.6$ ps ($\tau_L=12$ ps), which creates a slightly larger photoassociation window  ($R_{min}=84$ a$_0$, $R_{max}=117$ a$_0$), compared to ($R_{min}=85$ a$_0$, $R_{max}=110$ a$_0$) for the pulse with $\tau_C=34.8$ ps ($\tau_L=15$ ps). This increase of $P_{0_g^-}(E_0)$ has to be related to the increase of the probability density within the photoassociation window in the initial collisional state ($\int_{R_{min}}^{R_{max}}|\Psi_{\Sigma,E_0}(R')|^2dR'=0.00044$ and 0.00076, see also Ref.\cite{eluc04}, Fig.3), which is a signature of the nearly total population transfer in this range of R.

For the pulse with $\tau_C=34.8$ ps ($\tau_L=15$ ps) we have shown that the energy dependence  of $\bar{P}_{0_g^-}(E)$ is proportional to that of the square of the overlap integral $|<0_g^- \ v_0=98|^3\Sigma_u^+ \ E>|^2$ \cite{elianeanne}.
Taking into account the very similar dynamics induced by the two pulses with $\tau_C=34.8$ ps ($\tau_L=15$ ps) and $\tau_C=46.6$ ps ($\tau_L=12$ ps), it is reasonable to assume that the energy variation of  $\bar{P}_{0_g^-}(E)$ is the same for both pulses. Therefore, for the same temperature, both pulses correspond to the same value of the ratio ${\cal P}_{0_g^-}(T)/\bar{P}_{0_g^-}(E_0)$, and  the total probability per pump pulse $\tau_C=46.6$ ps ($\tau_L=12$ ps) that a given pair of atoms to be photoassociated into the $0_g^-$ state at a temperature $T=54$ $\mu K$ can be estimated to:
\begin{equation} 
Z{\cal P}_{0_g^-}(T=54 \ \mu K, \tau_C=46.6 \  ps )=0.0113.
\label{eq:pa3}
\end{equation}

\subsection{Total number of molecules photoassociated per pump pulse}
For a number of N atoms in a volume V, the number of pairs of atoms is 
$\frac{N(N-1)}{2} \approx \frac{N^2}{2}$. Taking into account the spin degeneracy of the $Cs(6^2S)$  atomic state, $d_A=2$, and of the initial electronic state  $d_{^3\Sigma_u}=3$, the total number of molecules photoassociated in the excited state $0_g^-$ per pump pulse is:
\begin{equation}
{\cal N}=\frac{N^2}{2} {\cal P}_{0_g^-}(T) \frac{ d_{^3\Sigma_u}}{d_A^2}
\end{equation}
For a trap of volume $V=10^{-3}$ cm$^3$, at the temperature $T=54$ $\mu K$ and with a density of atoms $N_A=N/V=10^{11}$ cm$^{-3}$, the partition function is $Q(T)=59.86 \times 10^{-10}$ a$_0^{-3}$$=40.4 \times 10^{15}$ cm$^{-3}$.  Then the number of molecules photoassociated per pump pulse is:
\begin{equation}
{\cal N}_{(\tau_C=34.8 ps)}=0.69, \ \ {\cal N}_{(\tau_C=46.6 ps)}=1.40
\end{equation}
For a repetition rate equal to $10^{8}$ Hz and supposing that each pulse acts on the same initial state, this gives $6.9 \times 10^{7}$ molecules per second for the pulse with $\tau_C=34.8$ ps ($\tau_L=15$ ps) and $1.4 \times 10^{8}$ molecules per second for the pulse with  $\tau_C=46.6$ ps ($\tau_L=12$ ps). 

The analysis of the energy-dependence of the photoassociation probability  $\bar{P}_{0_g^-}(E)$ for pulses leading to significant population transfer at large internuclear distances is in progress.

\section{Discussion : possible ways for optimization}
From the previous analysis, we may extract some directions on possible ways of optimizing the pulse. 
In the situation where the population transfer occurs mainly within the photoassociation window, for sufficiently large coupling $W_{max}$ the total adiabatic population transfer implies that the whole population for pair of atoms with relative distance lying in the  $[R_{min}, R_{max}]$ range, $P_{init} = \int_{R_{min}} ^{R_{max}}|\Psi_{g,E_0}(R',t=0)|^2| dR'$, is transferred to bound levels of the excited state and to the last bound levels of the ground state. The photoassociation yield can be optimized by designing the chirped pulse in order to maximize $P_{init}$. This can be achieved by increasing the photoassociation window with $\tau_C$ values as large as possible, under the condition $\hbar (\delta \omega + |\chi|  \tau_C )\le \delta_L^{at}$ so that only bound levels are populated. This yields an optimal value for $\tau_L$ which is $\tau_{opt} \sim \frac{8 \hbar \ln 2}{\delta_L^{at}}$. Since the condition (28) is fixing an upper value for $\tau_L$ to make focalization possible, we end with an upper limit for the detuning, and optimization can be achieved making use of the scaling laws governing the spectra of long range molecules. This will be further explored in future work, but for the detuning $\delta_L^{at}$= 2.652 cm$^{-1}$ considered in the present work, it is clear that the optimal pulse would be $\tau_L \sim$ 10 ps, corresponding to $P_{init}$= 1.27 $\times$ 10$^{-3}$, i.e. to a photoassociation rate increased by a factor 2.4 compared to the pulse with $\tau_C$= 47 ps ($\tau_L$=12 ps),  considered in the present work.\\
We have also obtained an increase of the population by a factor of 4.3 when considering the pulse  $\tau_C$=34.8 ps ($\tau_L$=15 ps),  and increasing the coupling by a factor of 16,  corresponding to a peak intensity $I_L$=3.36 MW cm$^{-2}$ for the transform limited pulse. \\
Finally,  a factor of 5 on the probability reported in the last column of Table I was obtained in case of the pulse with
$\tau_C$= 47 ps ($\tau_L$=12 ps), by reducing the detuning to 0.0695 cm$^{-1}$, therefore moving the value of $R_L$ to 150 a$_0$.
 
\section{Conclusion}
\label{sec:conclu}
We have investigated the possibilities offered by chirped laser pulses to optimize the yield of the photoassociation process. Following a previous paper \cite{eluc04}, time-dependent calculations have been presented for  the particular example of the  reaction Cs(6s) + Cs(6s) $^3 \Sigma_u^+ \to $ Cs$_2$ $ 0_g^-$ (6s + 6p$_{3/2}$)  involving ground state cesium atoms at a temperature $T \sim 54 \mu K$, colliding in presence of  laser pulses of different  spectral widths $\delta \omega$ with the same  linear chirp rate in the time domain $\chi$=-4.79 $\times$ 10 $^{-3}$ ps$^{-2}$ . The central frequency is red-detuned by $\delta_L \sim 2.65$ cm$^{-1}$ relative to the $D_2$ atomic  line, so  that the central frequency of the pulse excites at resonance the $v_0$=98 vibrational level in the outer well of the 0$_g^-$ level potential, mainly through  a vertical transition at the distance $R_L \sim$ 93.7 $a_0$ corresponding to  the outer classical turning point of the $v_0$ level.  The new aspects in the present work are :
\begin{itemize}
\item i) the calculations take into account the mixed state character of the initial collision state, described  by a statistical mixture of  stationary collision eigenstates representing thermal equilibrium at $T \sim 54 \mu K$. This choice reproduces correctly the spatial delocalization of the initial state, with  large de Broglie wavelength $\lambda_{DB}\sim$ 975 a$_0$. In the range of  detunings  considered here, experiments with a cw laser have demonstrated minima in the photoassociation rate corresponding to the nodes in the scattering wavefunctions : such nodes, common to the various eigenstates, are correctly reproduced in the present treatment. The continuum wavefunctions are represented as eigenstates in a large box (size $L_R \sim $19 500 a$_0$), using a mapped sine grid representation which involves a reasonable number of grid points. The time-dependent Schr\"odinger equation describing motion in two realistic potential curves coupled by the laser field is solved numerically. \\
 At ultracold temperatures, the threshold effects which govern the dependence of the photoassociation rate as a function of the energy $E$ have to be accounted for correctly. A high resolution analysis of such effects can be achieved by considering a large number of unity-normalized wavefunctions in the box, from which energy normalized functions are deduced. A proper estimate of the absolute value of the photoassociation rate has then been obtained from an incoherent average over a thermal distribution of the energy normalized wavefunctions.
\item ii) In the present paper, we have explored a large variety of pulses, all of them having the same central frequency, as described above, resonant with the level $v_0$=98. All of them have the same linear chirp parameter in the time domain, $\chi$, designed so that after the pulse, at time $t_p + T_{vib}/2$ (where $t_p$ corresponds to the maximum of the pulse, while $ T_{vib}$ is the classical vibrational period of $v_0$)  the vibrational wavepacket created in the excited state is focussing at the inner turning point. This choice is dictated by the objective of improving the efficiency of the stabilization step, where either by spontaneous or by induced emission the population is transferred to bound levels of the a$^3 \Sigma_u^+$ ground triplet state. This chirp rate can easily be deduced from the revival period, which can be defined provided the  levels populated stay in a small energy range around $v_0$. All the pulses are obtained from a gaussian transform-limited pulse, with the same peak intensity $I_L$= 120 kW/cm$^2$. They differ by the spectral width $ \delta \omega \sim (\tau_L)^{-1} $. By varying $\tau_L$ from 1 to 6 and 12 ps, various situations are analyzed, with narrow ($\sim$ 1 cm$^{-1}$) or broad   ($\sim$ 15 cm$^{-1}$) spectral width, long ($\tau_C \sim $  580 ps) or short ($\tau_C \sim$ 1 ps ) duration of the pulse stretched by chirping. Indeed, for a given choice  of $\delta \omega$ and $\chi$, two different pulses can be associated : the first one, with ($\tau_C / \tau_L) \approx 1$ is referred to as ``almost no-chirped'', the second one, with ($\tau_C/ \tau_L) \gg 1$ as ``really chirped''. The resonance window 2$\hbar |\chi| \tau_C$ explored by the instantaneous frequency during the time interval $[t_P - \tau_C, t_P + \tau_C]$ is larger in the second case.\\
Different situations are then encountered with respect to the validity of the impulsive approximation, or of the adiabaticity of the population inversion.     
\end{itemize}

 The conclusion of our study is that the ``really chirped'' pulses seem generally to be more efficient. In fact, we have observed two qualitatively different dynamical situations, depending upon the energy range defined by $\hbar (\delta \omega + |\chi|  \tau_C)$, i.e of the spectral width and the width of the resonance window : 
\begin{itemize}
\item Under the condition $\hbar (\delta \omega + |\chi|  \tau_C) < \delta_L^{at}$, already considered in previous work \cite{eluc04}, the concept of a photoassociation window is relevant. In spite of the delocalized character of the initial wavefunction, we observe after the pulse a wavepacket in the excited state localized around $R_L$ with a finite extension $[R_{min}, R_{max}]$, corresponding to the domain of variation of the outer turning points of the vibrational levels in the photoassociation window. Indeed, whereas during the time window $[t_P-\tau_C, t_P+\tau_C]$ levels outside this window are significantly populated, it is possible to optimize the parameter of the pulse so that an adiabatic model is valid where no population remains, after the pulse, outside the photoassociation window. By varying the size of the photoassociation window, it is possible to optimize the number of photoassociated molecules, and this direction should be further explored in future work.\\
For fixed detuning and spectral width, an increase of the laser coupling $W_L$ modifies the photoassociation dynamics. Having defined a parameter $\alpha$ such that adiabatic population transfer is taking place in the window $[-\alpha \tau_C, +\alpha \tau_C]$, we have shown that a larger intensity is increasing the time window, and therefore the width of the photoassociation window $[-\hbar\alpha |\chi|\tau_C, +\hbar\alpha |\chi|\tau_C]$, where total population inversion is taking place. Therefore as expected the photoassociation rate should be increased at large intensities.  However, the dynamics outside the photoassociation window at large internuclear distances ($R >$ 200 a$_0$) becomes less adiabatic. Rabi oscillations appear during the pulse, and levels outside the photoassociation window may remain populated after the pulse.
\item Another situation occurs when $\hbar (\delta \omega + |\chi|  \tau_C) > \delta_L^{at}$, since it is possible to transfer population to the continuum, or to highly excited vibrational levels in the excited potential $V_e(R)$. The results of the calculations indeed show  evidence for  population transfer at large internuclear distances. After the pulse, due to the attractive character of $V_e(R)$, with $-C_3/R^3$ asymptotic behaviour, the wavepacket created in the excited state is moving towards shorter distances $R$. For the uppermost levels, with an outer turning point located beyond $\sim$ 500 a$_0$, the vibrational half period $T_{vib}/2$ becomes comparable to the radiative lifetime, so that spontaneous emission may take  place before the intermediate distance region is reached, the photoassociated molecule decaying into a pair of atoms. In contrast, the wavepacket corresponding to lower levels has time to reach the intermediate region where spontaneous or induced emission may populate bound levels in the $V_g$ potential. The present calculations indicate that besides the acceleration due to the potential, the laser pulse has given a ``kick''   to the wavepacket in the $V_e$ curve, thus reducing the time necessary to reach the intermediate distance region where radiative stabilization of the molecule may take place. Optimization of the pulses in view of increasing this ``kick'' is a promising direction for future work, since at large distances   the continuum  wavefunctions representing the initial state display a  large amplitude  , and many vobrational levels of the $V_e$ potential are located close to dissociation limit, making the photoassociation process  very efficient at small detunings, while  it is well known from cw experiments that the bottleneck in such situation is the stabilization process \cite{masnou01}
\end{itemize} 

Both the present paper and Ref. \cite{eluc04} have been investigating the population transferred to the excited state due to  photoassociation with one chirped laser pulse. In order to get final conclusions more useful to experiments, the theoretical work should develop further in two  directions . First, besides the photoassociation step, future calculations should  investigate the efficiency of the stabilization process, bringing molecules to bound levels of the ground or lower triplet state, via spontaneous or induced emission. In particular the relevance of two-colour experiments where a second pulse with a larger central frequency is transferring population to the desired levels should be analyzed. Second, although the present estimation for photoassociation with a realistic repetition rate is promising, it does not take into account the fact that the second and further pulses are operating on a modified initial continuum state, where population has been extracted to be transferred to bound levels in the excited or ground state, or redistributed in the neighbouring continuum levels. The evolution of the atomic sample in the presence of a sequence of short pulses is an important issue, particularly in view of possible applications to condensates.

\section{Appendix A. The energy normalization of the ground state continuum wavefunctions calculated in a box.}
\label{sec:appA}

The $s$-wave ground state continuum wavefunctions (normalized in the energy scale: $<\Psi_{g,E}|\Psi_{g,E'} >=\delta(E-E')$) have the following asymptotic behaviour: 
\begin{equation}
|\Psi_{g,E}(R)|=\sqrt{\frac{2 \mu}{\pi \hbar^2}}\frac{\sin(k(R)+\eta_g)}{\sqrt{k(R)}}
\label{eq:asymptwf}
\end{equation}
where $\eta_g$ is a slowly varying phase and $k(R)$ the local wave number determined by the electronic ground state potential $V_g(R)$: $k(R)=1/\hbar$$\sqrt{2 \mu (E-V_g(R)) }$.
For $E=E_n$, the energy-normalized wavefunction $\Psi_{g,E_n}(R)$ is deduced from the unity-normalized wavefunction 
$\phi_{g,E_n}(R)$ using the density of states $\frac{\partial E}{\partial n}\mid_{E=E_n}$ at the energy $E_n$ \cite{landau}, \cite{ostrovsky01}. Indeed, the classically allowed domain for a continuum wavefunction is $R_t \le R \le L_R$, where $R_t$ is the inner turning point, and, in the standard semiclassical WKB form, the wavefunction can be written:
\begin{equation}
\Psi_{g,E=E_n}(R)=N_n\frac{1}{\sqrt{k(R)}} \sin \int_{R_t}^{R} (k(R')dR' + \frac{\pi}{4})
\end{equation} 
with the Bohr-Sommerfeld quantization:
\begin{equation}
 \int_{R_t}^{L_R}k(R')dR'=(n+\frac{1}{2})\pi
\end{equation} 
The normalization factor $(N_n)^{-2}$ is proportional to the density of states:
\begin{equation}
(N_n)^{-2}=\frac{2 \mu}{\pi \hbar^2}\frac{\partial E}{\partial n}\mid_{E=E_n}=\frac{1}{2} \int_{R_t}^{L_R}\frac{dR'}{k(R')}
\end{equation} 
Therefore: 
\begin{equation}
\Psi_{g,E=E_n}(R)=\lbrack \frac{\partial E}{\partial n}\mid_{E=E_n}\rbrack^{-1/2}\phi_{g,E_n}(R)
\label{eq:normen}
\end{equation}

\section{Appendix B. Chirp rate in the time domain for focussing the excited vibrational wavepacket at the inner turning point}

The gaussian linearly chirped laser pulse excites succesively several vibrational levels in the molecular surface $V_e$, creating a wavepacket located at the outer turning points of the corresponding vibrational wavefunctions. The center of the pulse excites at $t=t_P$ the level  $v_0$. We choose the negative value of the chirp parameter $\chi<0$ in the time domain such as the excited wavepacket, located initially at the outer turning points in the $0_g^-$ potential, to be focussed at the time $t = t_P + T_{vib}(v_0)/2$ at the inner turning points of the excited vibrational functions.  Then the chirp value has to compensate the dispersion in the vibrational periods $T_{vib}(v)$ of the vibrational levels $v$ resonantly excited at different times by the instaneous frequency $\omega(t)/2 \pi$ of the laser.

Let's note $E_v$ the binding energy of the level $v$ ($E_v>0$, 
$\frac{\partial E_v}{\partial v}<0$). The levels $v_0+1$, $v_0$  and  $v_0-1$ are resonantly excited at $t_P - (E_{v_0+1}-E_{v_0})/(\hbar \chi)<t_P $, $t_P$, and $t_P - (E_{v_0-1}-E_{v_0})/(\hbar \chi)>t_P $, reaching their inner turning points with the delays  $T_{vib}(v_0+1)/2$, $T_{vib}(v_0)/2$, and $T_{vib}(v_0-1)/2$, respectively. We deduce the value of the chirp $\chi$ from the condition that the components of the wavepacket to be in phase at the inner turning point. In a first approximation the $\chi$ is supplied by the condition:
\begin{eqnarray}
\frac{T_{vib}(v_0)}{2}= - \frac{E_{v_0+1}-E_{v_0}}{\hbar \chi}+ \frac{T_{vib}(v_0+1)}{2} \nonumber\\
= - \frac{E_{v_0-1}-E_{v_0}}{\hbar \chi}+ \frac{T_{vib}(v_0-1)}{2}
\label{eq:condfocus}
\end{eqnarray}
Using the definition of the vibrational period:
\begin{equation}
T_{vib}(v_0)=\frac{2 \pi \hbar}{|\frac{\partial E}{\partial v}|_{v_0}|}
\approx \frac{4\pi \hbar}{E_{v_0-1}-E_{v_0+1}}.
\label{eq:tvib}
\end{equation}
and of the revival period \cite{averbukh89}:
\begin{equation}
T_{rev}(v_0)= \frac{4 \pi \hbar}{|\frac{\partial^2 E}{\partial v^2}|_{v_0}|} \approx \frac{4 \pi \hbar}{ E_{v_0+1}+E_{v_0-1}-2E_{v_0}}.
\label{eq:trev}
\end{equation}
one can deduce the following value of $\chi$ from the Eqs. (\ref{eq:condfocus}):
\begin{equation} 
 \chi= -2 \pi \frac{T_{rev}(v_0)}{[T_{vib}(v_0)]^3}
\end{equation}

\section{Appendix C. Overlap of the initial continuum with the $0_g^-$ vibrational wavefunctions.}
\begin{figure}
\resizebox{0.5\textwidth}{!}{%
\includegraphics{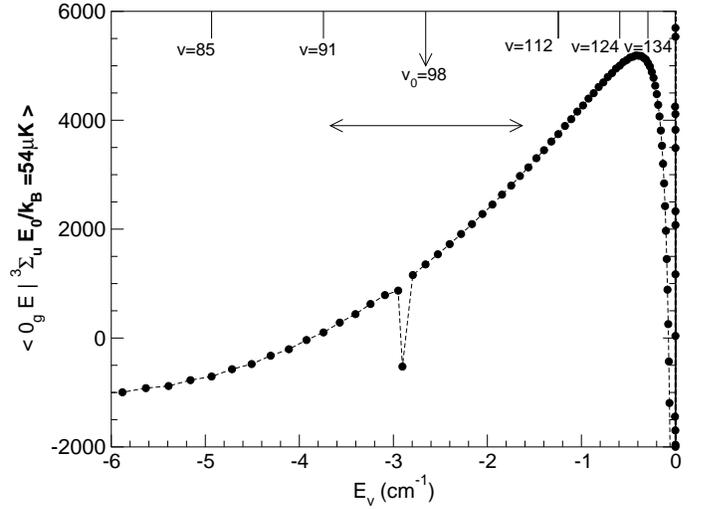}
}
\caption{The overlap between the initial stationary continuum state in the $a^3\Sigma_u^+$potential, corresponding to $E_0/k_B=54$ $\mu K$, and the wavefunctions of the bound levels of the $0_g^-$ potential, with energies $E_v$ between -6 and 0 cm$^{-1}$ under the 6s+6p$_{3/2}$ limit (the smaller value observed in the vicinity of -3 cm$^{-1}$ is due to the tunneling from the outer well to the inner well of the $0_g^-$ potential). The horizontal arrow shows the ``window'' of $0_g^-$  vibrational levels between v=92 and v=106, excited by the pulse studied in Ref. \cite{eluc04}.  }
\label{fig:recouvr}
\end{figure}

Fig. \ref{fig:recouvr}  represents the overlap integral $|<0_g^- \ E_v|^3\Sigma_u^+ \ E_0=k_B T>|$  between the initial collisional state in the $a^3\Sigma_u^+$potential, corresponding to $E_0/k_B=54$ $\mu K$, and the wavefunctions of the bound levels of the $0_g^-$ potential, with energies $E_v$ between -6 and 0 cm$^{-1}$ under the 6s+6p$_{3/2}$ limit.

{\bf Aknowledgements}

Discussions with Anne Crubellier and Ronnie Kosloff are gratefully acknowledged.
This work was performed in the framework of  the European Research Training Network ``Cold Molecules'', funded by the European Commission under contract HPRN CT 2002 00290. M.V. acknowledges for two three-months post-doctoral stays in Orsay funded by this contract.

\end{document}